\documentclass[epj]{svjour}
\usepackage{times}
\usepackage{graphics}
\newfont{\eufm}{eufm10}
\newfont{\cmu}{cmu10}

\begin{document}

\title{Beta-decay properties of $^{25}Si$ and $^{26}P$}

\author{J.-C. Thomas\inst{1}\inst{2} 
\and L. Achouri\inst{3} 
\and J. \"{A}yst\"{o}\inst{4} 
\and R. B\'eraud\inst{5} 
\and B. Blank\inst{1} 
\and G. Canchel\inst{1} 
\and S. Czajkowski\inst{1}
\and P. Dendooven\inst{6} 
\and A. Ensallem\inst{5} 
\and J. Giovinazzo\inst{1} 
\and N. Guillet\inst{1} 
\and J. Honkanen\inst{4} 
\and A. Jokinen\inst{4} 
\and A. Laird\inst{7}
\and M. Lewitowicz\inst{8} 
\and C. Longour\inst{9} 
\and F. de Oliveira Santos\inst{8} 
\and K. Per\"aj\"arvi\inst{10}
\and M. Stanoiu\inst{11}}

\institute{ Centre d'Etudes Nucl\'eaires de Bordeaux-Gradignan,
All\'ee du Haut-Vigneau, B.P. 120, 33170 Gradignan Cedex, France,
\and 
Instituut voor Kern- en Stralingsfysica, Celestijnenlaan 200D, 3001 Leuven, Belgium 
\and 
LPC Caen, 6 Boulevard du Mar\'echal Juin, 14050 Caen Cedex, France 
\and 
Department of Physics, PB 35 (YFL), 40014 University of Jyv\"{a}skyl\"{a}, Finland 
\and 
Institut de Physique Nucl\'eaire de Lyon, 69622 Villerbanne Cedex, France 
\and Kernfysisch Versneller Instituut, Zernikelaan 25, 9747 AA Groningen, Netherlands
\and 
Department of Physics and Astronomy, University of Edinburgh,
Edinburgh EH9 3JZ, United Kingdom 
\and Grand Acc\'el\'erateur
National d'Ions Lourds, B.P. 5027, 14076 Caen Cedex 5, France 
\and
Institut de Recherche Subatomique, 23 rue du Loess, B.P. 28, 67037
Strasbourg Cedex, France 
\and 
Lawrence Berkeley National Laboratory, 1 Cyclotron Road, Berkeley, CA 94720, USA
\and 
Institut de Physique Nucl\'eaire
d'Orsay, 15 rue G. Cl\'emenceau, 91406 Orsay Cedex, France}

\abstract{The $\beta$-decay properties of the neutron-deficient
nuclei $^{25}Si$ and $^{26}P$ have been investigated at the
GANIL/LISE3 facility by means of charged-particle and $\gamma$-ray
spectroscopy. The decay schemes obtained and the Gamow-Teller strength
distributions are compared to shell-model calculations based on
the USD interaction. $B(GT)$ values derived from the absolute
measurement of the $\beta$-decay branching ratios give rise to a
quenching factor of the Gamow-Teller strength of 0.6. A precise
half-life of $43.7\:(6)\:ms$ was determined for $^{26}P$, the
$\beta\!-\!(2)p$ decay mode of which is described.}

\PACS{ {29.30.Ep}{Charged-particle spectroscopy}\and {29.30.Kv}{X-
and gamma-ray spectroscopy}\and {23.90.+w}{} }

\maketitle

%%%%%%%%%%%%%%%%%%%%%%%%%%%%%%%%%%%%%%%%%%%%
%% MAINMATTER
%%%%%%%%%%%%%%%%%%%%%%%%%%%%%%%%%%%%%%%%%%%%

\section{Introduction}

\subsection{Generalities}

Over the last decades, $\beta$-decay properties of light
unstable nuclei have been extensively investigated in order to
probe their single-particle nuclear structure 
and to establish the proton and neutron
drip-lines. Hence, compilations of spectroscopic properties are
available for many $sd$ shell nuclei
\cite{[ti93],[ti95],[ti98],[en90],[en98]} from which 
nucleon-nucleon interactions were derived \cite{[usd]}. 
$\beta$-decay studies of nuclei having a large proton excess are
therefore useful to test the validity of these models when
they are applied to very unstable nuclei.

Moreover, in the standard $V\!\!-\!\!A$ description of 
$\beta$ decay, a direct link between experimental results
and fundamental constants of the weak interaction is 
given by the reduced transition
probability ${\cmu{f}}t$ of the individual allowed $\beta$ decays.
This parameter, which incorporates the phase space factor
${\cmu{f}}$ and the partial half-life $t\!=\!T_{1/2}/BR$
($T_{1/2}$ being the total half-life of the decaying nucleus and
$BR$ the branching ratio associated with the 
$\beta$ transition considered), can be written as follows:

\begin{equation}\label{ft}
{\cmu{f}}t\,=\,\frac{\mathcal{K}}{g_V^2\,|\!<f|\tau|i\!>\!|^2 +
g_A^2\: |\!<f|\sigma \tau|i\!>\!|^2}
\end{equation}
where $\mathcal{K}$ is a constant and where $g_V$ and $g_A$ are,
respectively, the vector and axial-vector current coupling
constants related to the Fermi and Gamow-Teller components of 
$\beta$ decay. $\tau$ and $\sigma$ are the isospin and the spin operators, 
respectively. Hence, the comparison of the measured
${\cmu{f}}t$ values and the computed Fermi and Gamow-Teller matrix
elements appears to be a good test of nuclear wave functions built
in the shell-model frame, stressing the role of the overlap
between initial and final nuclear states as well as the
configuration mixing occurring in parent and daughter states.
However, two systematic deviations from theoretical predictions
show the limitation of our theoretical understanding and treatment
of fundamental interactions. They are reported as the
\textit{mirror asymmetry anomaly in $\beta$ decay}
\cite{[nadya],[to73],[wi71],[wi70-2]} and the \textit{quenching of
the Gamow-Teller strength} \cite{[ax98],[ha92],[mu91]}.

\paragraph{Mirror asymmetry in $\beta$ decay:} 

\begin{figure}
\begin{center}
\resizebox{.48\textwidth}{!}{\includegraphics{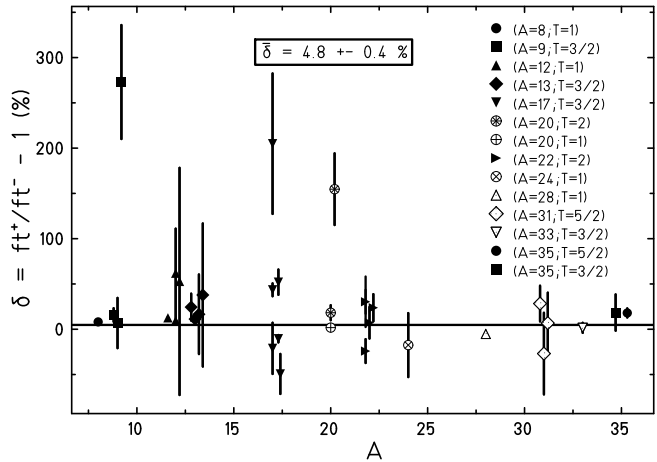}}
\caption[]{Systematics of the experimental values
   of the asymmetry parameter $\delta$ for nuclei with $A\!\le\!40$.
   Only allowed Gamow-Teller transitions with
   $log({\cmu{f}}t)\!\le\!6$ are considered.}
\label{jcdelta}
\end{center}
\end{figure}

This phenomenon is
related to the isospin non-conserving forces acting in the atomic
nucleus. If nuclear forces were charge independent, the $\beta^+$ ($EC$)
and the $\beta^-$ decays of analog states belonging
to mirror nuclei would be of equal strength. The deviation from
this simple picture is characterized by the asymmetry parameter
$\delta\,=\,({\cmu{f}}t^+/{\cmu{f}}t^-\:-\:1)$,
where the $+$ and $-$ signs are associated with the
decay of the proton- and the neutron-rich members of the mirror
pair, respectively. Figure~\ref{jcdelta} presents an updated
systematics of $\delta$ values measured for mirror nuclei with
$A\,\le\,40$. $39$ allowed Gamow-Teller mirror transitions with
$log({\cmu{f}}t)\!\le\!6$ pertaining to $14$ pairs of nuclei are
analyzed (see ref. \cite{[jct_thesis]} for details). They lead to
a mean deviation of about $5\,\%$ for these nuclei lying in the
$p$ and $sd$ shells. The asymmetry reaches $11\,(1)\,\%$ if only p
shell nuclei are considered, which stresses the interplay between the
Coulomb and the centrifugal barriers.

It was often attempted to explain the mirror asymmetry anomaly in
the $p$ shell either in terms of binding energy effects
\cite{[to73],[wi71],[ba92]} or by introducing the concept of "second
class currents" \cite{[wi70-1],[ku77],[wi00]}, which are not
allowed within the frame of the standard V-A model of the
weak interaction. None of the theoretical approaches were
able to reproduce the measured $\delta$ values.
Shell-model calculations are currently performed to test the isospin
non-conserving part of the interaction in $\beta$ decay
by studying the influence of isospin mixing effects and of radial 
overlap mismatches of nuclear wave functions on the Gamow-Teller 
matrix elements. These calculations are performed in the $p$ shell
and in the $sd$ shell, where reliable single-particle
nuclear wave functions are now available \cite{[nadya]}.

\paragraph{Gamow-Teller quenching:} 

The axial-vector coupling constant $g_A$ involved in
$\beta$ transitions of the Gamow-Teller type is not strictly constant and it
has to be renormalized in order to reproduce the $ft$ values
measured experimentally \cite{[bab85]}. The effective coupling constant
$g_{A,eff}\!=\!q \!*\!g_A$ is deduced
empirically from nuclear structure experiments and shows a 
slight variation over a wide range of masses: $q\!=\!0.820\,(15)$
in the $p$ shell \cite{[ch93]}, $q\!=\!0.77\,(2)$ in the $sd$
shell \cite{[wi83]} (giving a quenching factor $q^2$ of 0.6)
and $q\!=\!0.744\,(15)$ in the $pf$ shell \cite{[ma96]}.

Different theoretical approaches have been used in order to
derive the renormalization factor from core polarization effects
(due to particle-hole excitations), isobar currents and
meson exchange~\cite{[to83]}. Despite all
these efforts, the origin of the quenching effect is not very well
understood. Nevertheless, the Gamow-Teller strength function
$B(GT)$ $=(g_A/g_V)^2\,|\sigma\tau|^2$, which translates the
global response of the wave function to spin-isospin excitations
occurring in $\beta$ decay, is a useful link between
experimental results and theoretical predictions and it can be
used as a comparative tool.

\paragraph{Experimental development:} 
With the development of secondary radioactive beams and
other experimental techniques like the combination of helium-jet transport
systems with telescope detectors \cite{[se73],[ca84],[ro93]}, a
large set of neutron-deficient nuclei has been investigated since
the $\beta$-delayed proton emission was first observed forty years
ago \cite{[ba63]}. As $Q_{EC}$ values are increasing while
nuclei become more exotic, $\beta\!-\!p$ and $\beta\!-\!\gamma$
spectroscopic studies of neutron-deficient nuclei give the
opportunity to probe the Gamow-Teller strength function up to more
than 10 MeV in excitation energy. Hence, the whole energy window
open in $\beta$ decay can be covered both by spectroscopic
studies and charge exchange reactions \cite{[an91]}. Therefore,
the theoretical description of nuclear structure as
well as our understanding of the weak interaction can be tested
far from the stability line. As an illustration, we will report in
the following on the $\beta$-decay properties of two
neutron-deficient light nuclei, namely $^{25}Si$ and
$^{26}P$.

\subsection{Previous studies}

\subsubsection{Studies of $^{25}Si$} 

With a lifetime of 218~ms and
a $Q_{EC}$ value of about 13~MeV, the
$T_Z\!=\!-\frac{3}{2}$ nucleus $^{25}Si$ has been studied several times 
since the end of the 1960's. These previous studies will be
used in the present work to validate the analysis procedure implemented to
derive the $\beta$-decay properties of $^{26}P$. However, none of these 
studies measured the decay by $\gamma$ emission of excited states
fed in the $\beta$ decay of $^{25}$Si.

 The most recent $\beta$-delayed proton emission
study of $^{25}Si$ was performed by Robertson {\it et al.}
\cite{[ro93]}. It updates the first investigation of Reeder
{\it et al.} in 1966 \cite{[re66]}. In both experiments, the
individual proton group intensities were measured relative to the
most intense one, emitted by the isobaric analog state (IAS) in
$^{25}Al$. The absolute $\beta$-decay branching ratio of $12.2\,\%$
towards this state was derived from the associated
$log({\cmu{f}}t)$ value ($log({\cmu{f}}t)\!=\!3.28$), 
calculated assuming a pure Fermi
$\beta$ transition from the ground state of $^{25}Si$. It led to a
summed $\beta$ feeding of proton-unbound states of $^{25}Al$
equal to $38.1\,(15)\,\%$. This normalization
procedure is supported by the measurement of Hatori {\it et
al.} \cite{[ha92]}. In this work, absolute branching ratios for $\beta$
decay were determined by counting the total number of $\beta$ particles
emitted with the half-life of $^{25}$Si and the $\beta$ feeding of
the IAS in $^{25}Al$ was indeed found to be equal to
$14.6\,(6)\,\%$, giving rise to a $log({\cmu{f}}t)$ value of
$3.19\,(2)$. The summed feeding of the $^{25}Al$ proton-emitting
states was measured to be $40.7\,(14)\,\%$,
in good agreement with Robertson {\it et al.}

As mentioned above, in none of the experiments, the $\beta$-delayed
$\gamma$ decay of $^{25}Si$ was observed. As a consequence, the
$\beta$-decay branching ratios towards the proton-bound states of
$^{25}Al$ were tentatively estimated taking into account the
summed $\beta$ feeding and assuming that the relative
${\cmu{f}}t$ values of these states were equal to those of the
mirror states in $^{25}Mg$. The weak point of such a procedure is
that an average $\beta$ asymmetry of $20\,\%$ had to be taken into
account for all proton-bound state, which was assumed to be equally shared by the
proton-bound states disregarding their individual quantum
characteristics.

\subsubsection{Studies of $^{26}P$} 

Due to its $T_Z$ value of $-2$
and its short lifetime of less than $100\,ms$, $^{26}P$ has not
been investigated in detail so far. Compilations only report the
observation by Cable {\it et al.} of $\beta$-delayed
proton and two-proton emission from this nucleus
\cite{[ca84],[ca83]}. A half-life of $20^{+35}_{-15}\,ms$ was
deduced from the observation of the most intense proton group. It
led to a $\beta$ feeding of the IAS in $^{26}Si$ equal to
$1.9^{+3.5}_{-1.4}\,\%$ using a calculated $log({\cmu{f}}t)$ value
of $3.19$ (assuming a pure Fermi transition). Only three proton
groups were observed linking the IAS to the two lowest states of
$^{25}Al$ ($\beta\!-\!p$ decay) and to the ground state of
$^{24}Mg$ ($\beta\!-\!2p$ decay). The two decay modes of the
IAS were reported to be of similar magnitude. However, the
large $Q_{EC}$ value of $18\,MeV$ together with a proton separation 
energy of $5.5\,MeV$ for the daughter nucleus $^{26}Si$ 
are indications that
the $\beta$-delayed charged-particle spectrum may be rather
complex, involving a large number of proton groups.

\subsection{Present measurements} 

In our experiment, we determined the absolute branching ratios
for $^{25}$Si and $^{26}$P by relating the intensity of a given proton
or $\gamma$ line to the number of isotopes of each type implanted in 
our set-up. For $^{25}$Si, this measurement constitute a first unambigous
determination of branching ratios also for proton-bound levels. We will use
the decay of $^{25}$Si in part to test our analysis procedure, however, 
our study yields also new results for this nucleus, in particular for
the $\gamma$ decay of its $\beta$-decay daughter. In the case
of $^{26}$P, we deduce for the first time the feeding for other states
than the IAS and their decay by proton or $\gamma$ emission. 
Therefore, we could establish a complete decay scheme for
branches with more than about 1\% feeding for both nuclei for the first time. 

%%%%%%%%%%%%%%%%%%%%%%%%%%%%%%%%%%%%%%%%%%%%%%%%%%%%%%%%%%%%%%%%
%%%%%%%%%%%%%%%%%%%%%%%%%%%%%%%%%%%%%%%%%%%%%%%%%%%%%%%%%%%%%%%%
%%%%%%%%%%%%%%%%%%%%%%%%%%%%%%%%%%%%%%%%%%%%%%%%%%%%%%%%%%%%%%%%
\section{Experimental procedure}

\subsection{Fragment production and detection set-up}

In addition to $^{25}Si$ and $^{26}P$, the $\beta$-delayed proton
and two-proton emitters $^{22}Al$ \cite{[ac03]} and $^{27}S$
\cite{[ca01]} have been studied during the same experimental
campaign. The $\beta$-delayed proton emitter $^{21}Mg$
\cite{[se73]} and the $\beta$-delayed $\gamma$ emitter $^{24}Al$ 
\cite{[en90]} were also produced for calibration and efficiency
measurement purposes.

All nuclei have been produced in the fragmentation of a 95 MeV/u
${}^{36}$Ar$^{18+}$ primary beam with an intensity of about $2\mu{}Ae$ 
delivered by the coupled cyclotrons
of the GANIL facility. A $357.1\,mg/cm^2$ $^{12}C$ production
target was placed in the SISSI device \cite{[jo91]}, the high
angular acceptance and focusing properties of which increased the
selectivity of the fragment separation operated by the LISE3
spectrometer. The latter included a shaped $Be$ degrader 
(thickness 1062$\mu$m) at the
intermediate focal plane and a Wien filter at the end of the line
to refine the selection of the separated fragments.

\begin{figure}
\begin{center}
\resizebox{.48\textwidth}{!}{\includegraphics{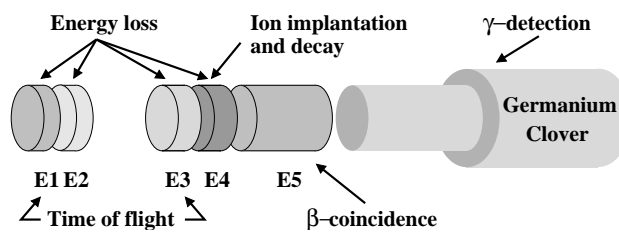}}
\caption[]{Schematic view of the identification 
   and detection set-up. It includes a germanium detector and five
   silicon detectors mounted in close geometry, where selected ions
   were identified by means of energy-loss and time-of-flight
   measurements. The last two detectors were used to observe the charged
   particles emitted in decay events in the implantation detector $E4$.}
\label{det_set-up}
\end{center}
\end{figure}

Ions of interest were implanted in the fourth element $E4$ of a silicon
stack (figure \ref{det_set-up}). The ion identification was
performed by means of time-of-flight and energy-loss measurements
with the silicon detectors $E1$ to $E4$ ($2*300\,\mu
m$ and $2*500\,\mu m$ in thickness, $4*600\,mm^2$ of surface). It led to
a precision in the counting rate of better than $1\,\%$ for $^{25}Si$
and about $3\,\%$ for the more exotic $^{26}P$ nucleus. The
production method in association with the high selectivity
of the LISE3 spectrometer gave rise to a very low contamination rate of
the selected species by only a few isotones (see table \ref{counting}).

\begin{table}
\begin{center}
\begin{tabular}{cccc}
\hline\noalign{\smallskip}
   nucleus & production & contamination & implanted ions\\
           & rate (pps) &  ($\%$)       &  ($*\,10^3$)\\
     \noalign{\smallskip}\hline\noalign{\smallskip}  
  $\:^{25}Si$ & $300$ & $\:\:<1$ & $\:492\,\:(\:\: 1)$ \\ 
  $\:\:^{26}P$ & $\:\:65$ & $\approx 13$ & $2180\,(70)$ \\
\noalign{\smallskip}\hline
\end{tabular}
\caption{Production rate, contamination and total number of
         selected ions during the experiment.} 
\label{counting}
\end{center}
\end{table}

Protons were detected in the implantation detector $E4$, in
coincidence with the observation of $\beta$ particles in the
detector $E5$ (with a thickness of $6\,mm$ and an area of  
$600\,mm^2$). A segmented germanium clover was finally
used to study the $\beta$-delayed $\gamma$ decay of implanted
ions.

%%%%%%%%%%%%%%%%%%%%%%%%%%%%%%%%%%%%%%%%%%%%%%%%%%%%%%%%%%%%%%%%
%%%%%%%%%%%%%%%%%%%%%%%%%%%%%%%%%%%%%%%%%%%%%%%%%%%%%%%%%%%%%%%%

\subsection{$\beta$-delayed proton spectroscopy}

Contrary to previous experiments \cite{[se73],[ca84],[ro93]} in
which ions were deposited at the surface of an ion catcher,
$\beta$-delayed protons are emitted inside the implantation
detector $E4$. As a first consequence, the proton spectrum rises
on a large $\beta$ background and the identification of
low-energy, low-intensity proton lines is difficult. Secondly, the
energy deposit in the detector $E4$ of an emitted proton cannot be
disentangled from the energy-loss contribution of the associated
$\beta$ particle and the recoiling ion.

\begin{figure}
\begin{center}
\resizebox{.4\textwidth}{!}{\includegraphics{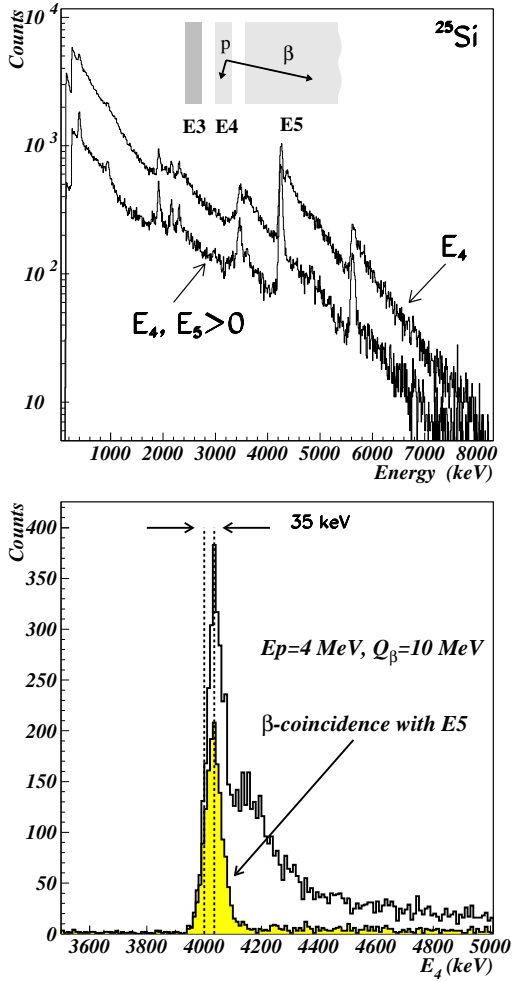}}
\caption[]{The upper part of the figure
    shows the influence of the
    $\beta$-coincidence condition ($E5\!>\!0$) on the energy
    spectrum delivered by the detector E4 for the setting on
    $^{25}Si$. The lower part of the figure presents a GEANT
    simulation of the effect of the coincidence condition on the 
    shape of a $\beta$-delayed proton emission 
    peak. The dotted lines show the $35 keV$ energy shift due to
    the $\beta$ pile-up, i.e. a $4 MeV$ proton peak is in fact observed
    at an energy of $4.035 MeV$. This shift depends on the implantation depth 
    and varies for the different nuclei studied in this work. 
    The coincidence condition does not alter
    this shift significantly.}
\label{coinc}
\end{center}
\end{figure}

To minimize these effects, ions were implanted in the last
$100\,\mu m$ of the detector $E4$ and a $\beta$ coincidence with
the thicker detector $E5$  
was requested in the analysis. As shown in the upper part of figure
\ref{coinc}, the $\beta$ particle energy deposit in the
coincidence spectrum was strongly reduced and proton peaks could
be easily identified and fitted with the help of Gaussian
distributions. The energy calibration of the detector $E4$ as well as the
measurement of the proton group intensities were performed on the
basis of this $E4$-$E5$ coincidence condition.

%%%%%%%%%%%%%%%%%%%%%%%%%%%%%%%%%%%%%%%%%%%%%%%%%%%%
\subsubsection{Energy calibration of the implantation detector}

The $\beta$ particle energy deposit leads to a
shift in energy of the Gaussian-like part of the proton peaks.
This effect could be reproduced by means of a GEANT simulation
\cite{[geant]}, as shown in the lower part of figure~\ref{coinc}
for a representative $\beta$-delayed proton peak. It could 
also be shown that the energy shift is roughly independent on the
proton and $\beta$-particle energies but linearly dependent on the
implantation depth of the ions, that is to say, on the
distance the $\beta$ particles travel in the detector $E4$ before
leaving it to enter
the coincidence detector $E5$. The energy calibrations of the
detector $E4$ for the settings on $^{21}Mg$, $^{25}Si$ and $^{26}P$
were therefore assumed to differ only by a shift proportional to
the implantation depths of the ions. 

The calibration parameters for the settings were deduced
from the identification of the major proton groups expected at
$1315\,(9)$, $1863\,(2)$, $2037\,(4)$, $2589\,(9)$, $4908\,(3)$
and $6542\,(3)$ $keV$  for the decay of $^{21}Mg$ \cite{[se73]}
and at $402\,(1)$, $1925\,(3)$, $2169$ $(7)$, $2312\,(4)$,
$3472\,(10)$, $4261\,(2)$ and $5630\,(2)\,keV$ for the decay of
$^{25}Si$ \cite{[ro93]}. The proton group energies were
recalculated using the excitation energies of the proton-emitting
states and the proton separation energies reported in a
compilation \cite{[en90]}.

%%%%%%%%%%%%%%%%%%%%%%%%%%%%%%%%%%%%%%%
\subsubsection{Proton detection efficiency}

Since ions were implanted at the end of the detector $E4$, the
proton detection efficiency $\mathcal{E}_p$ is very sensitive to
the implantation profile of the emitting ion and to the proton
energy. The detection efficiency for protons between $0.5$ and
$10\,MeV$ was computed by means of GEANT simulations. Following
experimental observations, implantation profiles were approximated
by Gaussian distributions in the beam direction (with a
standard deviation of $20\,\mu m$) and with a two
dimensional square shaped function in the orthogonal plane
\cite{[lo00]}. 

\begin{figure}
\begin{center}
\resizebox{.4\textwidth}{!}{\includegraphics{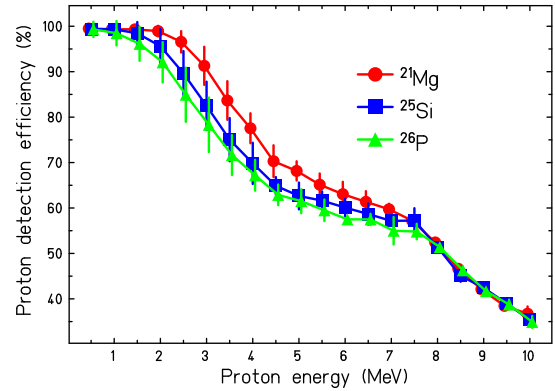}}
\caption[]{GEANT simulation of the proton detection
   efficiency of the implantation detector $E4$. The error bars on
   the plotted data are deduced from different parametrizations of the
   ion implantation profiles. An uncertainty due to the detector thickness
   is not included, as it is of the same order of magnitude as the 
   uncertainty of the implantation profile. (see text for details).}
\label{effp}
\end{center}
\end{figure}

Results are shown in figure \ref{effp}. An
uncertainty on the detection efficiency of less than $6~\%$ was
obtained. This uncertainty was determined by varying the 
implantation depth by $\pm$ 10 $\mu m$, which is roughly the width of the 
implantation distribution.

%%%%%%%%%%%%%%%%%%%%%%%%%%%%%%%%%%%%%
\subsubsection{Absolute intensities of the observed proton groups}

The absolute intensity $I_p^i$ of a given proton group $i$ was
derived from the following relation:

\begin{equation}
 I_p^i\,=\,\frac{Sc_p^i}{Kc_p*N_{impl}*\mathcal{E}_p^i}
\end{equation}
where $Sc_p^i$ is the area of the proton peak observed in the
coincidence spectrum ($E_5 > 0$), $Kc_p$ the normalization factor to be taken
into account due to the coincidence condition, $N_{impl}$ the
number of ions implanted in $E4$ and $\mathcal{E}_p^i$ the proton
detection efficiency for a given proton energy. 

\begin{figure}
\begin{center}
\resizebox{.48\textwidth}{!}{\includegraphics{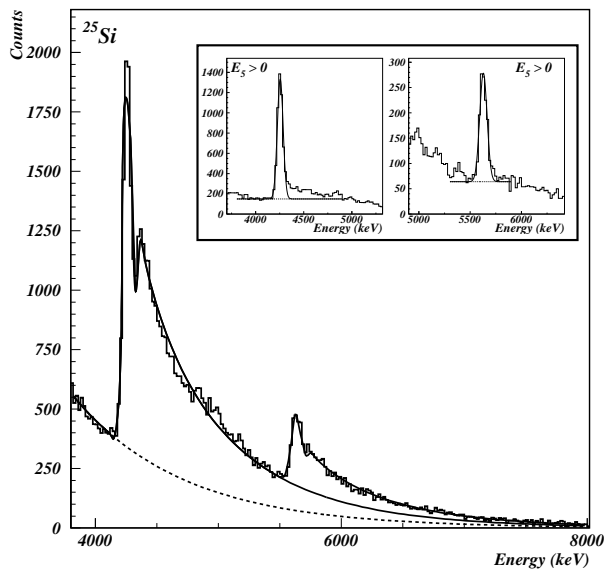}}
\caption[]{High energy part of the $\beta$-delayed
    proton spectrum in the decay of $^{25}Si$. The areas of the two
    main proton peaks above $4\,MeV$ obtained with (inset) and
    without (main figure) coincidence condition were used to extract
    an average normalization factor $Kc_p$ for the setting on
    $^{25}Si$.}
\label{pnorm}
\end{center}
\end{figure}

The extraction of the factor $Kc_p$ is
illustrated in figure \ref{pnorm} for the setting on $^{25}Si$.
Several proton peaks were fitted in the high
energy part of the $E4$ energy spectrum, where the
$\beta$ background is low enough and where proton peaks are well
separated. The $Kc_p$ coefficients were deduced from the average
ratio of the areas of the $\beta$-delayed proton peaks obtained
with and without coincidence condition. For the coincidence
spectrum, peaks were fitted by means of Gaussian distributions
on a linear background (see inserts of figure~\ref{pnorm}) leading to 
the $Sc_p^i$ values. For the unconditioned energy spectrum, fit 
functions convoluting a Gaussian distribution and an exponential tail 
on top of an exponential background were used. For each ion
of interest, the parameters of the exponential tail were fixed
regardless of the proton peak energies. The $Kc_p$ coefficients 
obtained were about $13\,\%$, with an uncertainty of $1$ to
$2\%$.

%%%%%%%%%%%%%%%%%%%%%%%%%%%%%%%%%%%%%%%%%%%%%%%
%%%%%%%%%%%%%%%%%%%%%%%%%%%%%%%%%%%%%%%%%%%%%%%%%%%%%%%%%%%%%%%%
\subsection{$\gamma$-ray spectroscopy}

As shown in figure \ref{det_set-up}, a segmented germanium clover detector
was placed at 0 degree, a few centimeters away from the silicon
stack. To reduce the dead time of the acquisition system, the
$\gamma$-ray signals were not used to trigger the data acquisition. As
a consequence, the probability to observe a $\gamma$ decay
depended on the type of radioactivity event that had triggered the
acquisition system. Since the energy loss of a proton in $E4$ was
larger than a few hundred $keV$, a trigger signal was
obtained each time a proton was emitted. Subsequent $\gamma$ rays
were then automatically detected, depending only on the Germanium
detector efficiency. 

On the other hand, most of the
$\beta$ particles emitted without accompanying protons did not lose
enough energy in $E4$ to trigger the acquisition system. As a
consequence, the trigger efficiency for $\beta$-$\gamma$ events was
given mainly by the fraction of the total solid angle under
which the large silicon detector $E5$ is seen from $E4$. This efficiency
was determined by means of $^{24}Al$ which decays by
$\beta$-$\gamma$ emission. The absolute intensities of the two
main $\gamma$ lines at $1077$ and $1369\,keV$ were measured and
compared with the expected values \cite {[en90]}. The
$\beta$-trigger rate was then derived, taking into account the
intrinsic efficiency of the Germanium detector, which was obtained
with conventional calibration sources. The overall
$\gamma$-detection efficiency in the $300$ to $2000\,keV$ range
was about $2$ to $3\,\%$, with a relative uncertainty of about
$20\,\%$. The $\beta$-trigger efficiency was equal to
$35.0\,(45)\,\%$.

To a large extent, corrections due to true summing
effects \cite{[de88]} were included in the calculated
$\beta$-trigger rate. However, this effect was not
under control when the acquisition was triggered by the detection
of protons, where the trigger efficiency was 100\%. 
Hence, $\gamma$-ray intensities could not be determined
reliably for $\beta\!-\!p-\!\gamma$ decay events and therefore
$\beta$-decay branching ratios towards proton-emitting states could not
be cross-checked by means of $\gamma$ spectroscopy.

%%%%%%%%%%%%%%%%%%%%%%%%%%%%%%%%%%%%%%%%%%%%%%%%%%%%%%%%%%%%%%%%
%%%%%%%%%%%%%%%%%%%%%%%%%%%%%%%%%%%%%%%%%%%%%%%%%%%%%%%%%%%%%%%%
%%%%%%%%%%%%%%%%%%%%%%%%%%%%%%%%%%%%%%%%%%%%%%%%%%%%%%%%%%%%%%%%

\section{Experimental results}

The $\beta$-decay properties of $^{25}Si$ are compared in the
following to the results obtained in previous work. For the two
settings on $^{25}Si$ and on $^{26}P$, the relative intensities of
the identified proton groups are given as well as the deduced
absolute $\beta$-decay branching ratios towards the proton-unbound
nuclear states of the daughter nuclei. The analysis of
$\beta$-delayed $\gamma$ spectra gives rise, for the first time,
to the measurement of the absolute feeding of the proton-bound
states. The decay schemes are then proposed and compared to
calculations performed in the full $sd$ shell by Brown
\cite{[bab]} with the OXBASH code \cite{[oxbash]} using the
USD interaction \cite{[usd]}. Finally, the Gamow-Teller strength
distributions are compared to those expected from the mirror
$\beta$ decays and to those extracted from the calculated $log({\cmu{f}}t)$ values.
The main characteristics of $^{26}P$ are given, including a
precise measurement of its lifetime as well as a derivation of its proton
separation energy $S_p$ and of its atomic mass excess $\Delta\,(^{26}P)$.

%%%%%%%%%%%%%%%%%%%%%%%%%%%%%%%%%%%%%%%%%%%%%%%%%
%%%%%%%%%%%%%%%%%%%%%%%%%%%%%%%%%%%%%%%%%%%%%%%%%%%%%%%%%%%%%%%%
\subsection{$\beta$-decay study of $^{25}Si$}

%%%%%%%%%%%%%%%%%%%%%%%%%%%%%%%%%%%%%%%%%%%%%%%%%%%%%%%%%%%%%%%%
\subsubsection{$\beta$-delayed proton emission}

\begin{figure}
\begin{center}
\resizebox{.48\textwidth}{!}{\includegraphics{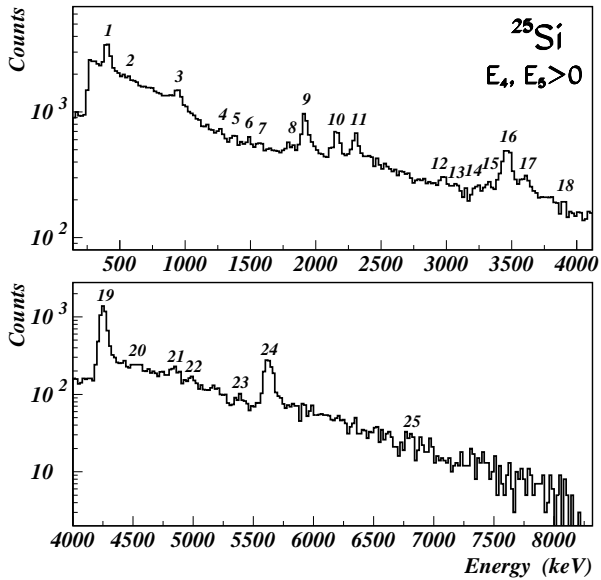}}
\caption[]{$\beta$-delayed proton spectrum obtained 
      in coincidence with a signal in $E5$ in the decay of $^{25}Si$.
      The peak labels correspond to the peak numbers used in 
      tables~\ref{tab:si25} and \ref{25sifeeding}.}
\label{25siprot}
\end{center}
\end{figure}

The $\beta$-delayed proton emission spectrum obtained in
coincidence with the detector $E5$ is shown in figure
\ref{25siprot} for the setting on $^{25}Si$.
Most of the proton groups reported in previous work by 
Robertson {\it et al.} \cite{[ro93]} and Hatori {\it et al.}
\cite{[ha92]} have been identified. Their center of mass energies
and their relative intensities are compared in table
\ref{tab:si25} and discussed in the following.

{\it Missing transitions}: Nine of the thirty-two proton groups
reported by Robertson {\it et al.} were not observed in the
present work. Six of these transitions (see table~\ref{tab:si25})
were already not observed
in the work of Hatori {\it et al.} and it is therefore
plausible that they are due to the decay of $\beta$-p contaminants
in the experiment performed by Robertson {\it et al.}. The
three remaining missing transitions have a relative intensity
lower than $1\,\%$ and it may be that the residual
$\beta$ background in the spectrum conditioned by $E5$  is too
large in the present experiment to allow for their
identification.

{\it Identification of the observed proton groups}: All
the observed proton groups were attributed to proton transitions
reported in the work of Robertson {\it et al.} on the basis
of their measured center of mass energies. Two groups at
$2980\,(9)$ and $3899\,(2)\,keV$ were tentatively identified as being the 
same transitions as those at $3021\,(9)$ and $3864\,(20)\,keV$ by Robertson 
{\it et al.} although the
energy differences are about $40\,keV$. The proton group at
$401\,(1)$ in the present work corresponds most likely to the 
$382\,(20)\,keV$ group of Robertson 
{\it et al.} because of their high relative intensities. The transition at
$1377\,(6)\,keV$ was attributed to an emission
from the IAS of $^{25}Al$ and was therefore identified as the
transition at $1396\,(20)\,keV$ of Robertson {\it et al}. 
The same is most likely true for the transitions at $1573\,(7)$
and $1592\,(20)\,keV$. The proton group at
$3326\,(6)\,keV$ was observed at the same energy as in the work of
Hatori {\it et al.} and corresponds most likely to the transition at
$3342\,(15)\,keV$ in reference \cite{[ro93]}. All other
transitions identified were measured at energies differing by less
than $15\,keV$ with respect to the work of Robertson {\it et al.}

{\it High-energy proton groups}: Only one of the three
high energy transitions reported by Zhou {\it et al.}
\cite{[zh85]} was identified at $6802\,(7)\,keV$. Its relative
intensity of $2.2\,(5)\,\%$ is significantly higher that the
values given in references \cite{[ha92]} and \cite{[zh85]},
which might reflect an underestimation of the proton detection
efficiency at high energies in the present work.

{\it New transition}: A new proton transition at $3077\,(14)\,keV$
(label 13) was 
observed but could not be attributed. 
Due to its low intensity of $0.25\,(11)\,\%$,
the transition could not be assigned neither by means of
a $\beta\!-\!p\!-\!\gamma$ coincidence nor by any other means.

{\it Assignment of proton transitions}: Apart from the transition
at $3326\,keV$ which, according to Hatori {\it et al.}, originates 
from the $5597\,keV$ excited level in $^{25}Al$, all identified
proton groups were assigned following the work of Robertson
{\it et al.} The deduced energies and absolute $\beta$-decay
branching ratios of the
proton-unbound states of $^{25}Al$ are presented in table
\ref{25sifeeding}. The obtained excitation energies are compared
to the data of the compilation \cite{[en98]}. Large discrepancies
of more than $25\,keV$ are found for the proton groups at
$2980$, $3899$ and $5407\,keV$. The IAS of $^{25}Al$ was found
at an excitation energy of $7892\,(2)\,keV$, in agreement with the
value of $7896\,(6)$ reported by Robertson {\it et al.}~\cite{[ro93]}.

\renewcommand{\arraystretch}{.95}
\begin{table*}
\begin{center}
\begin{tabular}{|c|r|r|r||c|r|r||r|r|} \hline
\multicolumn{4}{|c||}{This experiment} & \multicolumn{3}{c||}{Robertson \textit{et al.} \cite{[ro93]}}&
\multicolumn{2}{c|}{Hatori \textit{et al.} \cite{[ha92]}}\\
\hline
                           & C.M. Energy &Relative $\;\;\;$&Absolute $\;\;\;$&
                           & C.M. Energy &Relative $\;\;\;$& C.M. Energy &Relative $\;\;\;$\\
{\raisebox{2.ex}[0pt]{Peak}}&($keV$)$\;\;\;$&intensity
($\%$)&intensity
($\%$)&{\raisebox{2.ex}[0pt]{Peak}}&($keV$)$\;\;\;$&intensity
($\%$)&($keV$)$\;\;\;$&intensity ($\%$)\\\hline
  1  &$ 401\:(\:\:1)$&$ 49.8\:(48)$  &$ 4.75\:(32)$&  1  &$ 382  \:(20)$&$ 73.7  \:(3)$&$ 403  \:(\:\:1)$&$ 57.7  \:(63)$\\  \hline
  2  &$ 555\:(11)$    &$  7.2\:(27)$  &$ 0.69\:(25)$&  2 &$   550  \:(25)$&$  <2.5  \:(1)$&\multicolumn{2}{c|}{     }     \\  \hline
  3  &$ 943\:(\:\:2)$&$ 17.1\:(24)$  &$ 1.63\:(20)$&  3  &$   943.4\:(-)$ &$  17     \:(1)$&$   945\:(\:\:2)$ &$  11.5     \:(24)$\\  \hline
                         \multicolumn{4}{|c||}{     }&  4  &$1040\:(20)$ &$  1.53\:(3)$&\multicolumn{2}{c|}{     }     \\
                         \cline{1-7}
  4  &$1268\:(\:\:6)$&$  6.1\:(18)$  &$ 0.58\:(17)$&  5  &$1272  \:(20)$&$        2.26 \:(7)$&\multicolumn{2}{c|}{     }     \\
  \cline{1-7}
  5  &$1377\:(\:\:6)$&$  4.3\:(12)$  &$ 0.41\:(11)$&  6  &$1396  \:(20)$&$        2.89 \:(5)$&\multicolumn{2}{c|}{     }     \\
  \cline{1-7}
  6  &$1489\:(\:\:7)$&$  5.0\:(15)$  &$ 0.48\:(14)$&  7  &$1501  \:(20)$&$        2.90 \:(5)$&\multicolumn{2}{c|}{     }     \\  \hline
  7  &$1573\:(\:\:7)$&$  4.3\:(13)$  &$ 0.41\:(12)$&  8  &$1592  \:(20)$&$        1.46 \:(3)$&$1586  \:(\:\:3)$&$        2.1 \:(\:\:9)$\\  \hline
                         \multicolumn{4}{|c||}{     }&  9  &$1685\:(20)$ &$  0.93     \:(6)$&\multicolumn{2}{c|}{     }     \\  \hline
  8  &$1804\:(\:\:8)$&$  6.1\:(14)$  &$ 0.58\:(13)$& 10  &$1805  \:(15)$&$        6.73 \:(6)$&$1791  \:(\:\:3)$&$        5.4 \:(10)$\\  \hline
  9  &$1917\:(\:\:2)$&$ 23.5\:(27)$  &$ 2.24\:(21)$& 11  &$1925.2\:(-)$ &$    27.4  \:(1)$&$1925\:(\:\:3)$ &$    17.6  \:(13)$\\  \hline
 10  &$2162\:(\:\:4)$&$ 18.1\:(26)$  &$ 1.73\:(22)$& 12  &$2165  \:(10)$&$    17.2  \:(1)$&$2169  \:(\:\:7)$&$    12.8  \:(11)$\\  \hline
 11  &$2307\:(\:\:4)$&$ 16.5\:(25)$  &$ 1.57\:(21)$& 13  &$2311.4\:(-)$ &$    14.1  \:(1)$&$2312\:(\:\:4)$ &$    11.2  \:(\:\:9)$\\  \hline
                         \multicolumn{4}{|c||}{     }& 14  &$2373\:(20)$ &$  2.02     \:(3)$&\multicolumn{2}{c|}{     }     \\
                         \cline{5-7}
                         \multicolumn{4}{|c||}{     }& 15  &$2453\:(25)$ &$  0.40     \:(2)$&\multicolumn{2}{c|}{     }     \\
                         \cline{5-9}
                         \multicolumn{4}{|c||}{     }& 16  &$2486\:(25)$ &$  0.96     \:(2)$&$2483\:(\:?\:)$ &$  <1.4\quad\quad     $\\
                         \cline{5-9}
                         \multicolumn{4}{|c||}{     }& 17  &$2608\:(25)$ &$  0.39     \:(5)$&$2636\:(10)$ &$  0.5     \:(\:\:2)$\\  \hline
 12  &$2980\:(\:\:9)$&$  1.7\:(\:\:7)$  &$ 0.16\:(\:\:7)$& 18  &$3021  \:(15)$&$      3.74 \:(9)$&$3022  \:(\:\:9)$&$      5.0 \:(14)$\\  \hline
 13  &$3077\:(14)$    &$  2.6\:(12)$  &$ 0.25\:(11)$& \multicolumn{3}{|c||}{     }&\multicolumn{2}{c|}{     }     \\  \hline
 14  &$3231\:(\:\:8)$&$  5.4\:(13)$  &$ 0.51\:(12)$& 19  &$3237  \:(15)$&$    4.15 \:(5)$&$3243  \:(10)$&$    2.4 \:(\:\:6)$\\  \hline
 15  &$3326\:(\:\:6)$&$  5.9\:(12)$  &$ 0.56\:(11)$& 20  &$3342  \:(15)$&$       6.57 \:(6)$&$3356  \:(30)$&\\
 \cline{1-8}
 16  &$3463\:(\:\:3)$&$ 28.1\:(34)$  &$ 2.68\:(26)$& 21  &$3466  \:(10)$&$    34.5  \:(1)$&$3472  \:(10)$&{\raisebox{1.5ex}[0pt]{$    44.7  \:(48)$}}\\  \hline
 17  &$3610\:(11)$    &$  5.9\:(18)$  &$ 0.56\:(17)$& 22  &$3597  \:(10)$&$    10.86 \:(8)$&$3608  \:(\:\:5)$&$    13.3 \:(18)$\\  \hline
 18  &$3899\:(\:\:2)$&$3.4\:(\:\:7)$&$ 0.32\:(\:\: 6)$& 23  &$3864  \:(20)$&$   1.15 \:(7)$&$3852  \:(\:\:8)$&$   3.9 \:(12)$\\  \hline
 19  &$4252\:(\:\:2)$&$100\:(10)$  &$ 9.54\:(66)$& 24  &$4258.3\:(-)$ &$100    \:(2)$&$4261\:(\:\:2)$ &$100\quad\quad  $\\  \hline
                         \multicolumn{4}{|c||}{     }& 25  &$4303\:(20)$ &$  3.32     \:(7)$&\multicolumn{2}{c|}{     }     \\  \hline
 20  &$4545\:(10)$    &$  6.6\:(18)$  &$ 0.63\:(17)$& 26  &$4556  \:(20)$&$     1.28 \:(5)$&$4552  \:(\:\:8)$&$     1.3 \:(\:\:6)$\\  \hline
                         \multicolumn{4}{|c||}{     }&  27  &$4626\:(25)$ &$  0.25     \:(1)$&$4612\:(10)$ &$  0.4    \:(\:\:4)$\\  \hline
 21  &$4850\:(\:\:6)$&$ 10.3\:(17)$  &$ 0.98\:(15)$& 28  &$4853  \:(15)$&$    7.29 \:(7)$&$4841  \:(\:\:5)$&$    16.7 \:(18)$\\  \hline
 22  &$4986\:(\:\:8)$&$ <4.9\:(\:\:9)$&$<0.47\:(\:\:8)$& 29  &$4992  \:(15)$&$     2.30 \:(4)$&$4977  \:(\:\:5)$&$     1.4 \:(\:\:4)$\\  \hline
 23  &$5407\:(\:\:7)$&$3.6\:(\:\:7)$&$ 0.34\:(\:\:6)$& 30  &$5394  \:(20)$&$  1.98 \:(5)$&$5366  \:(\:\:6)$&$  0.8 \:(\:\:3)$\\  \hline
                         \multicolumn{4}{|c||}{     }&  31  &$5549\:(15)$ &$  3.19     \:(6)$&\multicolumn{2}{c|}{     }     \\  \hline
 24  &$5624\:(\:\:3)$&$ 25.1\:(27)$  &$ 2.39\:(20)$    & 32  &$5630  \:(10)$&$ 16.9  \:(2)$&$5630  \:(\:\:2)$&$ 21.1  \:(15)$\\  \hline\hline
 25  &$6802\:(\:\:7)$&$2.2\:(\:\:5)$&$ 0.21\:(\:\:4)$ &ref.~\cite{[zh85]} & $6520  \:(10)$&$ 0.72  \:(4)$ & $6795  \:(17)$&$ 0.7  \:(\:\:3)$\\  \hline
\end{tabular}
\caption{$\beta$-delayed proton emission of $^{25}Si$. The center
  of mass energy and the relative intensity of the identified proton
  groups are compared to previous experimental data. The relative and absolute
  intensities of the $\beta$-delayed proton transitions obtained in
  this work are also reported.}
\label{tab:si25}
\end{center}
\end{table*}

The overall agreement between the three $\beta$-delayed proton
decay studies of $^{25}Si$ is reasonable, leading to a summed
$\beta$-decay branching ratio towards the proton unbound states of
$^{25}Al$ equal to $35\,(2)\,\%$ (this work), $38\,(2)\,\%$
\cite{[ro93]} and $41\,(1)\,\%$ \cite{[ha92]}. The difference
originates for a large part from the determination of the absolute
intensity of the least energetic proton group at
about $400\,keV$ in the center of mass. This proton transition
is reported in the previous work to compete with a $\gamma$
deexcitation of the associated nuclear state, but no evidence was
found in the $\gamma$-decay spectrum for such a decay mode.

\renewcommand{\arraystretch}{1.2}
\begin{table*}
\begin{center}
\begin{tabular}{|l l l l|r||r||r|} \hline
\multicolumn{4}{|c|} {C.M. proton energies ($keV$) in the decay to
$^{24}Mg$ states}&\multicolumn{3}{|c|}{Excitation energies and
$\beta$ feeding}
\\ \multicolumn{4}{|c|}{    }
& \multicolumn{3}{|c|}{(B.R.) of $^{25}Al$ proton-unbound states}
\\ \cline{5-7} Ground state   &$\;\; 1369\:keV$ &$\;\;
4123\:keV$           &$\;\; 4238\:keV$          &This work& ref.
\cite{[en98]}& B.R. ($\%$) \\\hline $\:1\!-\! 401\,(1)$& & &
&$2672\,( 1)$&$2673.5\,( 6)$&$ 4.8\,(3)$\\  \hline
$\:7\!-\!1573\,(7)$& & & &$3844\,( 7)$&$3858.8\,( 8)$&$
0.4\,(1)$\\ \hline $\:9\!-\!1917\,(2)$&$\!\!\!\! 2\!-\!
555\,(11)$& & &$4189\,( 2)$&$4196  \,( 3)$&$ 2.9\,(3)$\\  \hline
$11\!-\!2307\,(4)$&$\!\!\!\! 3\!-\! 943\,( 2)$& & &$4582\,(
2)$&$4583 \,( 4)$&$ 3.2\,(3)$\\ \hline
                  &$\!\!\!\! 4\!-\!1268\,( 6)$&                           &                          &$4908\,( 6)$&$4906  \,( 4)$&$ 0.6\,(2)$\\  \hline
$15\!-\!3326\,(6)$&                           & & &$5597\,(
6)$&$5597  \,( 5)$&$ 0.56\,(11)$\\  \hline
                  &$\!\!\!\!10\!-\!2162\,( 4)$&                           &                          &$5802\,( 4)$&$5808  \,( 6)$&$ 1.7\,(2)$\\  \hline
$18\!-\!3899\,(2)$&                           & & &$6170\,(
2)$&$6122 \,( 3)$&$ 0.32\,(6)$\\  \hline
                  &$\!\!\!\!12\!-\!2980\,( 9)$&                           &                          &$6620\,( 9)$&$6645  \,( 4)$&$ 0.16\,(7)$\\  \hline
                  &$\!\!\!\!14\!-\!3231\,( 8)$&                           &                          &$6871\,( 8)$&$6881  \,( 6)$&$ 0.5\,(1)$\\  \hline
$21\!-\!4850\,(6)$&$\!\!\!\!16\!-\!3463\,( 3)$& & &$7107\,(
3)$&$7121 \,( 6)$&$ 3.7\,(2)$\\  \hline
$22\!-\!4986\,(8)$&$\!\!\!\!17\!-\!3610\,(11)$& & &$7255\,(
7)$&$7240 \,( 3)$&$ <\,1.0\,(6)$\\  \hline $23\!-\!5407\,(7)$& & &
&$7678\,( 7)$&$7637  \,( 6)$&$ 0.34\,(6)$\\ \hline
$24\!-\!5624\,(3)$&$\!\!\!\!19\!-\!4252\,( 2)$&$\!\!\!\!
6\!-\!1489\,( 7)$&$\!\!\!\!5\!-\!1377\,( 6)$&$7892\,( 2)$&$7902
\,( 2)$&$12.8\,(8)$\\  \hline
                  &$\!\!\!\!20\!-\!4545\,(10)$&$\!\!\!\! 8\!-\!1804\,( 8)$&                          &$8193\,( 6)$&$8186  \,( 3)$&$ 1.2\,(2)$\\  \hline
$25\!-\!6802\,(7)$&                           & & &$9073\,(
7)$&$9065 \,(10)$&$ 0.21\,(4)$\\  \hline
\end{tabular}
\caption{Excitation energies and $\beta$ feeding of $^{25}Al$
         proton-unbound nuclear states. Absolute branching ratios for the 
         present work, which are deduced from the absolute intensity 
         measurements of the
         $\beta$-delayed proton transitions, are given in the last column.}
\label{25sifeeding} 
\end{center}
\end{table*}

Regarding the absolute $\beta$ feeding of the IAS in $^{25}Al$,
the value of $12.8\,(8)\,\%$ obtained in this work is in good
agreement with the theoretically expected value of $12.2\,\%$ used
by Robertson {\it et al.} and is significantly lower than the one measured
by Hatori {\it et al}. It leads to a $log({\cmu{f}}t)$ value
of $3.25\,(3)$ for the $\beta$ decay of $^{25}Si$ towards the
IAS in $^{25}Al$. This result confirms the assumption that the
involved $\beta$ transition is almost purely of the Fermi type, since a
$log({\cmu{f}}t)$ value of $3.28$ is expected in this case
\cite{[ro93]}.

%%%%%%%%%%%%%%%%%%%%%%%%%%%%%%%%%%%%%%%%%%%%%%%%%%%%%%%%%%%%%%%%
\subsubsection{$\beta$-delayed $\gamma$ decay}

\begin{figure}[hht]
\begin{center}
\resizebox{.48\textwidth}{!}{\includegraphics{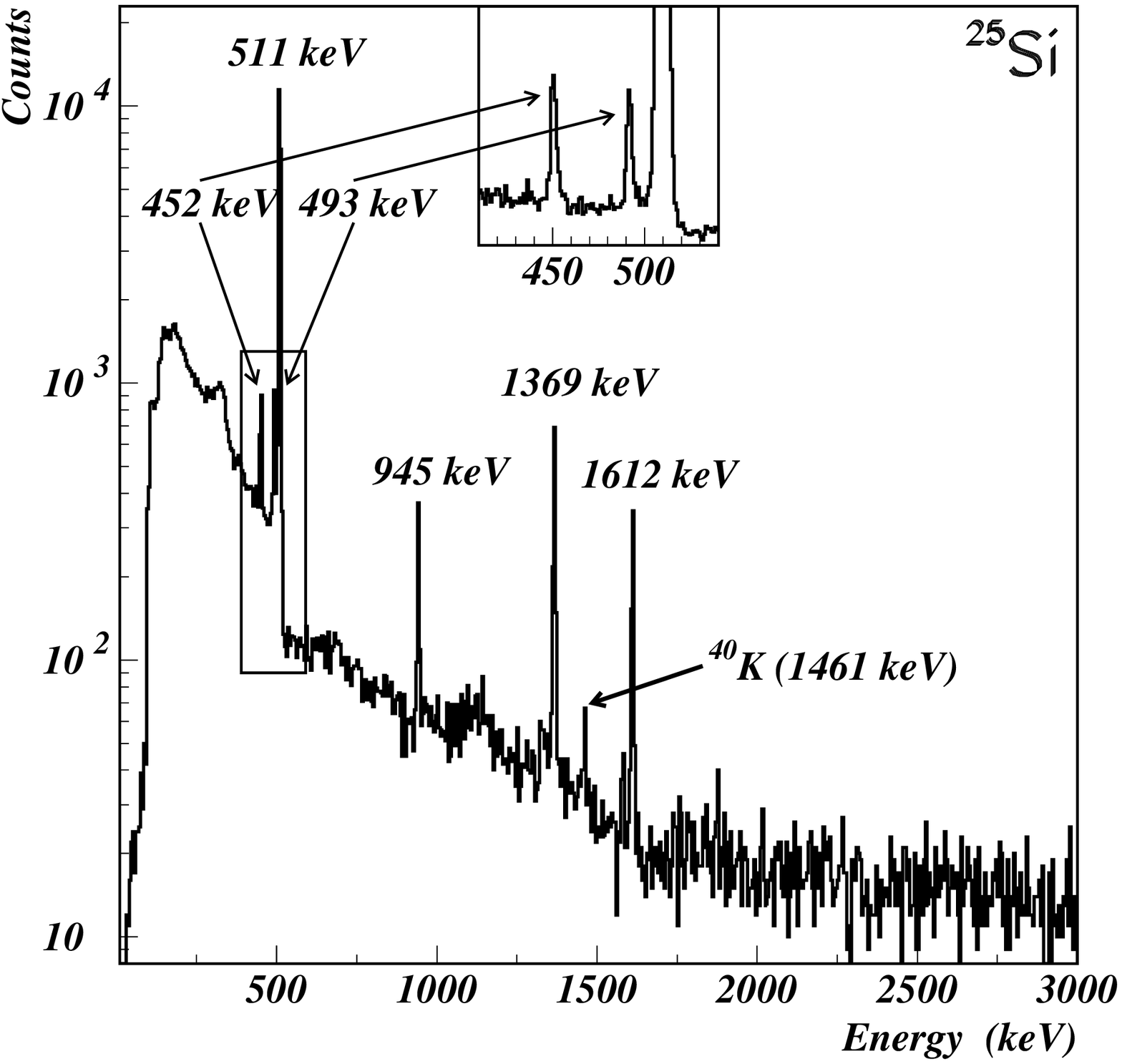}}
\caption[]{$^{25}Si$
      $\gamma$-decay spectrum. All $\gamma$ lines except the one at 1461~$keV$
      are attributed to the decay of $^{25}Si$.}
\label{25sigam}
\end{center}
\end{figure}

The $\gamma$-ray spectrum obtained in the decay of $^{25}Si$ is shown
in figure \ref{25sigam}. The four $\gamma$ lines at $452$
(absolute branching ratio of $18.4\,(42)\,\%$), $493$ ($15.3\,(34)\,\%$), $945$
($10.4\,(23)\,\%$) and $1612\,keV$ ($15.2\,(32)\,\%$) were
assigned to the $\beta$-delayed $\gamma$ decay of $^{25}Si$. The
last $\gamma$ line is a doublet of two $\gamma$
rays from the decay of the $\frac{7}{2}_1^+$ states at 
$1612.4\,keV$ in $^{25}Al$ and at $1611.7\,keV$ in its
daughter nucleus $^{25}Mg$ \cite{[en90]}. Taking into account the
expected contribution of this second transition in $^{25}Mg$, 
the absolute intensity of
the $1612\,keV$ $\gamma$ ray in $^{25}$Al was deduced to be equal to
$14.7\,(32)\,\%$.

The $\gamma$ lines at $493$ and $945\,keV$ are associated with the
deexcitation of the $\frac{3}{2}_1^+$ state at $945\,keV$ in
$^{25}Al$ towards its $\frac{5}{2}_1^+$ ground state and towards
the $\frac{1}{2}_1^+$ excited state at $452\,\,keV$. The intensity
ratio of the two lines
$I_\gamma(945)/I_\gamma(493)\,=\,68\,(26)\,\%$ is in agreement
with the value of $79\,(6)\,\%$ obtained in an in-beam
experiment \cite{[en90]}.

Since the intensities of the $493$ and
$452\,\,keV$ $\gamma$ rays were found to be equal within their
uncertainties, we conclude that the $452\,\,keV$ state is
not fed directly in the $\beta$ decay of $^{25}Si$. Such a
$\beta$ transition would be indeed a first-forbidden one and is
therefore unlikely to be observed in the present experiment.

No $\gamma$ rays were observed at $845$, $1338$ and $1790\,keV$.
Therefore, it was assumed that the $\frac{7}{2}_2^+$ proton-bound state
of $^{25}Al$ at $1790\,keV$ \cite{[en98]} is not fed in the
$\beta$ decay of $^{25}Si$. Hence, the measurements of the
absolute intensities of the three $\gamma$ lines at $493$, $945$
and $1612\,keV$ led to a summed $\beta$-decay branching ratio towards
the proton-bound excited states of $^{25}Al$ of $41\,(5)\,\%$ (see
table \ref{25sifeeding2} for details). Taking into account the 
previously determined summed $\beta$-decay branching ratio
towards the proton-unbound states ($35\,(2)\,\%$), this leads to
an absolute $\beta$ feeding of the $^{25}Al$ ground state of
$24\,(7)\,\%$.\\

The $\gamma$ line observed at $1369\,keV$ corresponds to the
deexcitation of the first excited state of $^{24}Mg$  populated in the
$\beta\!-\!p$ decay of $^{25}Si$. Due to the quite low
$\gamma$-detection efficiency and the weakness of most of the
proton transitions feeding excited states of $^{24}Mg$,
neither the $4_1^+\!\rightarrow\!2_1^+$ nor the
$2_2^+\!\rightarrow\!4_1^+$ transitions were seen. Only a few
counts at an $E4$ energy of about $4.25\,MeV$ were observed in
coincidence with the $\gamma$ ray at $1369\,keV$, in agreement with the
assignment
of the strongest proton group to the IAS in $^{25}Al$. 
The $\gamma$ line at $1461\,keV$ is the well-known background $\gamma$ ray from $^{40}K$.

%%%%%%%%%%%%%%%%%%%%%%%%%%%%%%%%%%%%%%%%%%%%%%%%%%%%%%%%%%%%%%%%
\subsubsection{$\beta$-decay scheme of $^{25}Si$}

\begin{figure*}
\begin{center}
\resizebox{.80\textwidth}{!}{\includegraphics{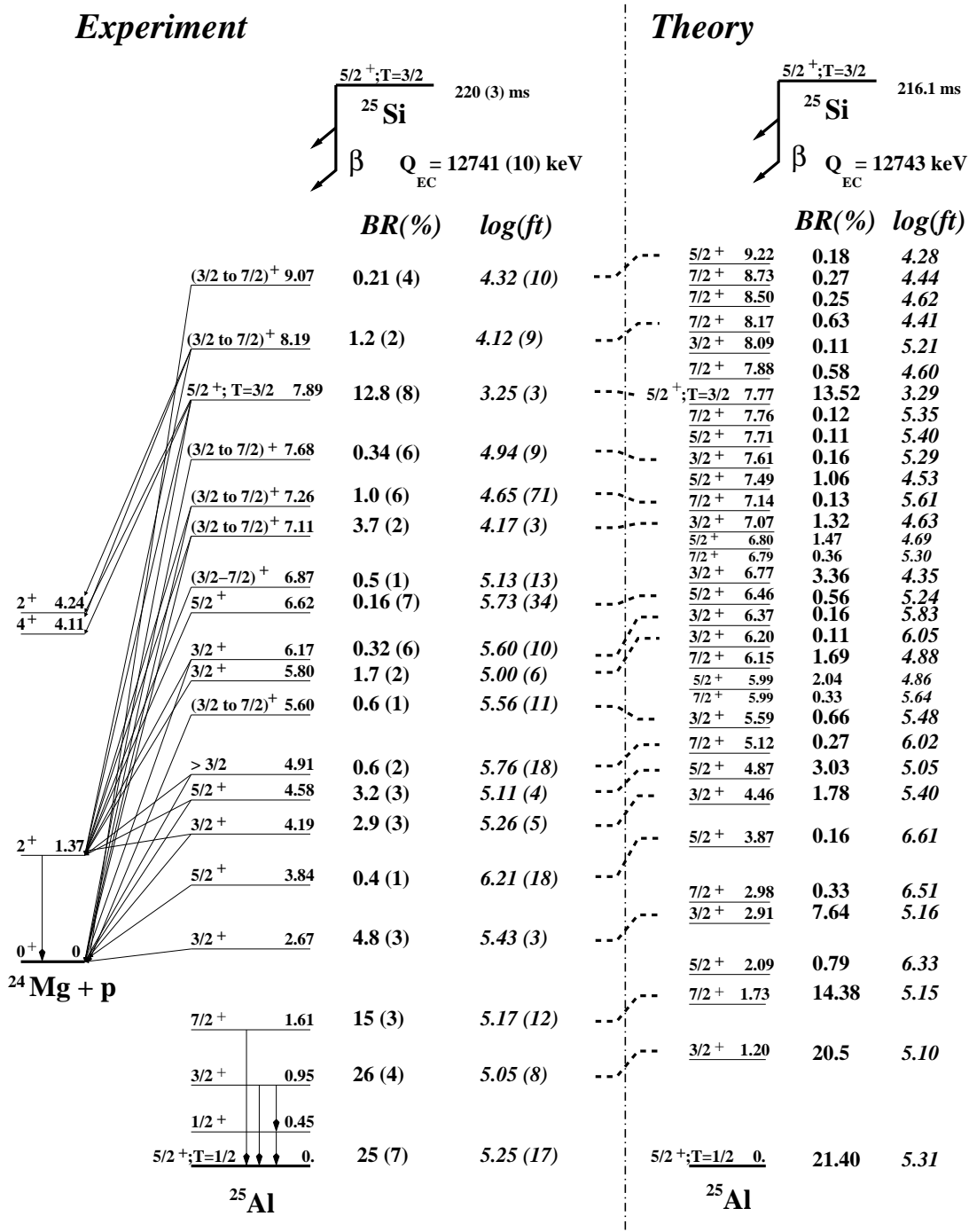}}
\caption[]{$\beta$-decay scheme of $^{25}Si$. 
   Experimental results are compared to shell-model
   calculations performed by Brown \cite{[bab]} in the full $sd$
   shell with the USD interaction and the OXBASH code.
   The dotted lines tentatively connect experimentally determined levels
   to levels predicted by theory.}
\label{25sitrans}
\end{center}
\end{figure*}

Figure \ref{25sitrans} shows the $\beta$-decay scheme proposed
for $^{25}Si$. The experimental branching ratios and the corresponding
$log({\cmu{f}}t)$ values are compared to shell-model
calculations performed by Brown \cite{[bab]}. Only
excited states predicted to be fed with a branching ratio of more than
$0.1\,\%$ are taken into account.
In terms of nuclear structure,
the agreement between experimental results and theoretical
calculations appears to be very good, most of the observed nuclear
states being reproduced by the model within a few hundred 
$keV$.

\begin{figure}
\begin{center}
\resizebox{.48\textwidth}{!}{\includegraphics{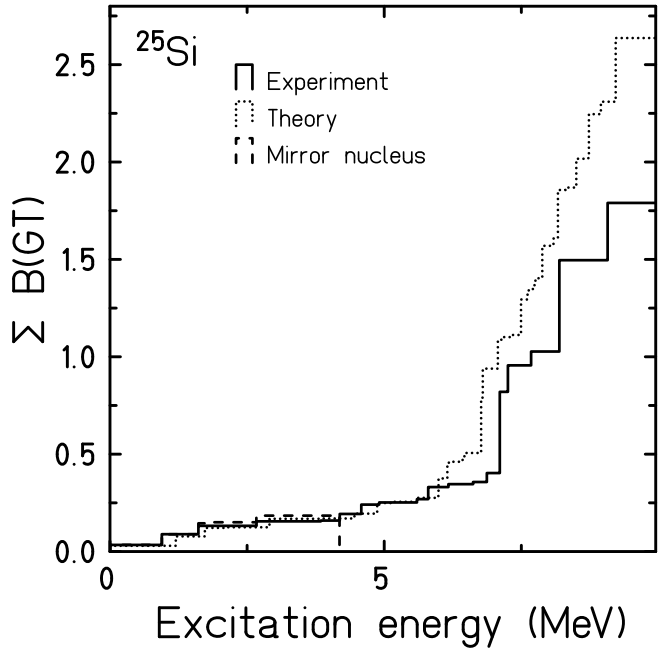}}
\caption[]{Theoretical and
   experimental distributions of the summed Gamow-Teller strength
   ($\Sigma B(GT)$) for the decay of $^{25}$Si. At low excitation energy, 
   the distributions are also compared to the one obtained from the mirror
   $\beta$ decay of $^{25}$Na assuming isospin symmetry. 
   The error of the Gamow-Teller
   strength distribution as determined in the present work is about 20\%.}
\label{25bgt}
\end{center}
\end{figure}

The summed Gamow-Teller strength distribution as a
function of the excitation energy of $^{25}Al$ is shown
in figure \ref{25bgt}. The experimental distribution is in good
agreement with the one deduced from the shell-model calculations
up to $6\,MeV$. Beyond, the model predicts the feeding of a lot of
high-energy excited states by low intensity $\beta$ transitions
that are not visible experimentally. Due to the small phase-space
factor ${\cmu{f}}$ associated with such transitions, the related
$B(GT)$ values are of importance, which explains the divergence at
more than $6\,MeV$ of excitation energy. The global agreement below
$6\,MeV$ is obtained for $11$ individual $\beta$ transitions for
which the Gamow-Teller strength is quenched equivalent to a quenching 
factor of about 0.6.

\renewcommand{\arraystretch}{1.5}
\begin{table}
\begin{center}
\begin{tabular}{|c|r|r|} \hline
\multicolumn{1}{|c|}{states populated in $^{25}Al$ }&\multicolumn{1}{|c|}{Excitation energy ($keV$)}& B.R.
($\%$)\\\hline
 $\frac{5}{2}^+_1$   &$   0        $&$25\,(7)$\\
$\frac{3}{2}^+_1$&$ 944.8 \:(4)$&$26\:(4)$\\
 $\frac{7}{2}^+_1$&$1612.4
\:(4)$&$15\:(3)$\\ \hline
\end{tabular}
\caption{$\beta$-decay branching ratios towards proton-bound nuclear
states of $^{25}Al$.} 
\label{25sifeeding2}
\end{center}
\end{table}

At low excitation energy, the
Gamow-Teller strength seems to be close to the one expected from
the $\beta$ decay of the $^{25}Si$ mirror nucleus, assuming that
nuclear forces are isospin independent ($\delta\!=\!0$).
Unfortunately, the error for the $\beta$-decay branching ratios towards
these states is too large (see table~\ref{tab:asymm} below) due to the uncertainty on the
$\gamma$-detection efficiency and the individual values of the
asymmetry parameter $\delta$ could not be derived precisely for
the $(A\!=\!25,T\!=\!3/2)$ isospin multiplet
($^{25}Na$,$^{25}Mg$, $^{25}Al$,$^{25}Si$).

%%%%%%%%%%%%%%%%%%%%%%%%%%%%%%%%%%%%%%%%%%%%%%%%%%%%%%%%%%%%%%%%
%%%%%%%%%%%%%%%%%%%%%%%%%%%%%%%%%%%%%%%%%%%%%%%%%%%%%%%%%%%%%%%%
\subsection{$\beta$-decay study of $^{26}P$}

\begin{figure}
\begin{center}
\resizebox{.50\textwidth}{!}{\includegraphics{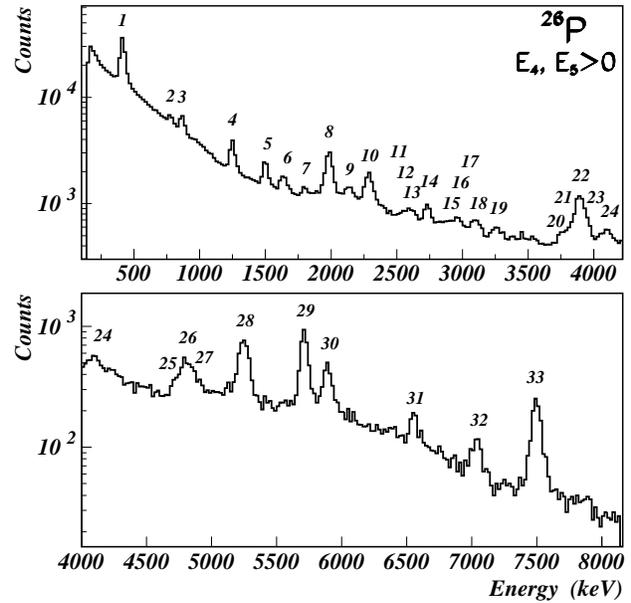}}
\caption[]{$\beta$-delayed one-proton and
    two-proton spectrum obtained in coincidence with a signal in $E5$ in the
    decay of $^{26}P$. The peak labels correspond to the peak numbers used in 
      table~\protect\ref{tab:26pprot}.}
\label{26pprot}
\end{center}
\end{figure}

The experimental procedure established and tested with
$^{25}$Si is now applied to $^{26}$P.
The $\beta$-delayed proton spectrum conditioned by the detection
of $\beta$ particles in $E5$ is shown in figure \ref{26pprot}. The
contamination from other $\beta$-delayed proton emitters 
was determined from energy-loss and
time-of-flight measurements to be $0.6\,\%$ for $^{27}S$ and
$1.2\,\%$ for $^{25}Si$.
The most intense proton transition occurring in the decay of the
latter nucleus is expected at
$4.25\,MeV$. It was not observed in the present setting and all
identified proton groups were therefore attributed to the decay of
$^{26}P$.

\renewcommand{\arraystretch}{1.}
\begin{table*}
\begin{center}
\begin{tabular}{|c|r|r|r||l|} \hline
Peak & C.M. Energy   &Relative &Absolute& $\gamma$ rays observed\\
     &($keV$)$\;\;\;$&intensity ($\%$)&intensity ($\%$)&in coincidence    \\  \hline
  1  &$ 412\,(\:\:2)$&$100.0\,(71)$    &$17.96\,(90) $  &                 \\  \cline{1-4}
  2  &$ 778\,(\:\:3)$&$  4.3\,(\:\:5)$ &$ 0.78\,(\:\:7)$&                 \\  \cline{1-4}
  3  &$ 866\,(\:\:2)$&$  9.5\,(10)$    &$ 1.71\,(15) $  &                 \\  \cline{1-4}
  4  &$1248\,(\:\:2)$&$  8.4\,(\:\:8)$ &$ 1.51\,(12) $  &                 \\  \hline
  5  &$1499\,(\:\:2)$&$  5.5\,(\:\:5)$ &$ 0.99\,(\:\:7)$&$493,\,945$      \\  \hline
  6  &$1638\,(\:\:3)$&$  3.6\,(\:\:4)$ &$ 0.65\,(\:\:6)$&$452$            \\  \hline
  7  &$1798\,(\:\:4)$&$  1.1\,(\:\:3)$ &$ 0.20\,(\:\:5)$&$452,\,493$      \\  \hline
  8  &$1983\,(\:\:2)$&$ 13.3\,(11)$    &$ 2.39\,(16) $  &                 \\  \hline
  9  &$2139\,(\:\:4)$&$  3.0\,(\:\:8)$ &$ 0.54\,(14) $  &$452,\,493,\,1338$\\  \hline
 10  &$2288\,(\:\:3)$&$  8.2\,(\:\:9)$ &$ 1.47\,(12) $  &$1612$            \\  \hline
 11  &$2541\,(\:\:6)$&$  0.5\,(\:\:2)$ &$ 0.09\,(\:\:3)$&                  \\  \cline{1-4}
 12  &$2593\,(13)$   &$  1.5\,(\:\:3)$ &$ 0.27\,(\:\:6)$&                  \\  \cline{1-4}
 13  &$2638\,(18)$   &$  0.6\,(\:\:2)$ &$ 0.11\,(\:\:4)$&                  \\  \cline{1-4}
 14  &$2732\,(\:\:4)$&$  2.6\,(\:\:4)$ &$ 0.47\,(\:\:6)$&                  \\  \cline{1-4}
 15  &$2855\,(17)$   &$<\,0.8\,(\:\:2)$&$<\,0.14\,(\:\:4)$&                \\  \hline
 16  &$2908\,(11)$   &$  0.3\,(\:\:3)$ &$ 0.06\,(\:\:5)$&$452,\,493$       \\  \hline
 17  &$2968\,(\:\:5)$&$  1.8\,(\:\:3)$ &$ 0.32\,(\:\:5)$&$452,\,493$       \\  \hline
 18  &$3097\,(\:\:6)$&$  1.7\,(\:\:4)$ &$ 0.31\,(\:\:6)$&$452,493,845,1790$\\  \hline
 19  &$3258\,(\:\:4)$&$  1.9\,(\:\:2)$ &$ 0.23\,(\:\:4)$&$452,\,493$       \\  \hline
 20  &$3766\,(\:\:9)$&$  2.0\,(\:\:4)$ &$ 0.36\,(\:\:7)$&$452$             \\  \hline
 21  &$3817\,(\:\:6)$&$  0.7\,(\:\:3)$ &$ 0.13\,(\:\:5)$&$452,\,945$       \\  \hline
 22  &$3879\,(\:\:3)$&$  4.4\,(\:\:6)$ &$ 0.79\,(12) $&$1369$              \\  \hline
 23  &$3920\,(\:\:5)$&$  6.7\,(\:\:9)$ &$ 1.21\,(14) $&                    \\  \cline{1-4}
 24  &$4097\,(\:\:5)$&$<\,2.1\,(\:\:3)$&$<\,0.37\,(\:\:4)$&                \\  \hline
 25  &$4719\,(\:\:6)$&$  1.3\,(\:\:2)$ &$ 0.24\,(\:\:4)$&$452$             \\  \hline
 26  &$4793\,(\:\:3)$&$  3.0\,(\:\:4)$ &$ 0.54\,(\:\:6)$&                  \\  \hline
 27  &$4858\,(\:\:4)$&$  2.5\,(\:\:3)$ &$ 0.44\,(\:\:5)$&$452$             \\  \hline
 28  &$5247\,(\:\:3)$&$  7.6\,(13)$    &$ 1.37\,(22) $&                    \\  \hline
 29  &$5710\,(\:\:3)$&$  7.8\,(\:\:7)$ &$ 1.40\,(11) $&$452,\,845,\,1790$  \\  \hline
 30  &$5893\,(\:\:4)$&$  4.1\,(\:\:8)$ &$ 0.73\,(13) $&$1612$              \\  \hline
 31  &$6551\,(\:\:4)$&$  1.2\,(\:\:5)$ &$ 0.21\,(\:\:8)$&$452,\,945$       \\  \hline
 32  &$7039\,(\:\:5)$&$  1.0\,(\:\:1)$ &$ 0.17\,(\:\:2)$&                  \\  \cline{1-4}
 33  &$7494\,(\:\:4)$&$  3.4\,(\:\:3)$ &$ 0.61\,(\:\:5)$&                  \\  \hline
\end{tabular}
\caption{$\beta$-delayed one-proton and two-proton emission of $^{26}P$.
The center of mass energy, the relative intensity and the absolute
intensity of proton groups identified in figure \ref{26pprot} are
given. The last column reports the $\gamma$ rays observed in
coincidence with the proton peaks. Transitions 22 and 28 are due to
two-proton emission from the IAS of $^{26}$Si.}
\label{tab:26pprot}
\end{center}
\end{table*}

%%%%%%%%%%%%%%%%%%%%%%%%%%%%%%%%%%%%%%%%%%%%%%%%%%%%%%%%%%%%%%%
\subsubsection{$\beta$-delayed proton emission}

The center of mass energies as well as the relative and absolute intensities
of the identified $\beta$-delayed proton or two-proton transitions
are given in table \ref{tab:26pprot}. The large amount of produced
nuclei allowed to performed $\beta\!-\!p\!-\!\gamma$ coincidences.
Table  \ref{tab:26pprot} indicates the energy of the $\gamma$ rays
that were seen in coincidence with the corresponding proton peaks.
All $\gamma$ lines except the one at $1369\,keV$ are assigned to
the decay of excited states of $^{25}Al$. The $\gamma$ ray at
$1369\,keV$ is due to the $\beta$-delayed two-proton decay
(transition $22$, see below) of $^{26}P$ towards the first
excited state of $^{24}Mg$.

{\it Proton and two proton emission of the IAS in $^{26}Si$}: 
By means of energy
considerations and $\beta\!-\!p\!-\!\gamma$ coincidences, the five
transitions labelled $29$ to $33$ could be assigned to the proton
decay of the isobaric analog state in $^{26}Si$. Its excitation
energy was determined to be equal to $13015\,(4)\,keV$. Based on
this first set of assignments, the two groups $22$ and $28$
were identified as two-proton transitions from the IAS to the
first excited state and to the ground state of $^{24}Mg$. The
IAS excitation energy deduced from these two transitions is
slightly higher in energy ($13036\,keV$ instead of $13015\,keV$) due to
the lower pulse-height defect in two-proton emission. In the case of 
nearly back-to-back emission of the two protons, the recoil has an energy 
close to zero, whereas for parallel emission of the two protons, 
the recoil has a maxiumun energy comparable to one-proton emission.
As the process of two-proton emission is supposed to be isotropic,
the average recoil energy is about half compared to the one-proton 
emission value leading to a lower pulse height defect.

As described later, the determination of the excitation energy
of the IAS was used to calculate the atomic mass excess of
$^{26}P$. The precise measurement of the $^{26}P$ lifetime (see
below), and the summed $\beta$-decay branching ratio of
$5.28\,(35)\,\%$ towards the IAS in $^{26}Si$ led to a
$log({\cmu{f}}t)$ value for this state of $3.13\,(5)$. This result
is close to the expected model-independent value of $3.186$.

{\it Emission from known excited states of $^{26}Si$}: 
Due to their energy, transitions $3$,
$8$, $12$ and $13$ were attributed to the decay of the previously
observed excited states of $^{26}Si$ at $6350\,(25)$, $7489\,(15)$,
$8570\,(30)$ and $8120\,(20)\,keV$ \cite{[en90]}. Although
transition $12$ is therefore expected to populate the first
excited state of $^{25}Al$, no $\gamma$ ray at $452\,keV$ was
observed in coincidence and the assignment of the transition is
somewhat questionable. However, the intensity of this 
proton peak is rather weak which prevents most likely the observation 
of a coincident $\gamma$ ray.

{\it $\beta\!-\!p\!-\!\gamma$ coincidences}:
$\gamma$ rays occurring in the deexcitation of $^{25}Al$
states were observed in coincidence with the proton groups
labelled $5\!-\!7$, $9\!-\!10$, $16\!-\!21$, $25$ and $27$. Therefore, 
these transitions were assigned to transitions between initial and final 
states based on $p-\gamma$ coincidences. No $\gamma$ rays were seen in coincidence with the transitions $1$, $2$ and $4$, although they are quite 
intense. Hence, they were assumed to populate directly the ground state of
$^{25}Al$.

{\it Assignments based on energy criteria}: The
excitation energies of the proton-emitting states in $^{26}Si$
derived from $\beta\!-\!p\!-\!\gamma$ coincidences were used to
identify the proton groups $14$, $23$ and $26$ as transitions
from these states towards the ground state of $^{25}Al$ (c.f. table
\ref{26pfeeding}).

{\it Unassigned proton transitions}: The three transitions $11$,
$15$ and $24$ could not be assigned to the decay of excited states
of $^{26}Si$. Neither a coincident $\gamma$ ray could be observed 
for these proton lines, nor
their energy corresponds to an energy difference of identified levels.
The 3 transitions represent less than $1.11\,\%$ (one-sigma limit) of the
measured $\beta$-decay strength of $^{26}P$.

{\it Atomic mass excess $\Delta(^{26}P)$}: The mass excess of $^{26}P$ 
was derived from the following relation:

\begin{equation}
\displaystyle \Delta (^{26}P)\,=\,\displaystyle \Delta (^{26}Si) +
E^*(IAS) + \Delta E_c - \Delta_{nH}\; \displaystyle
\end{equation}
where $\Delta (^{26}Si)$ is the atomic mass excess of $^{26}Si$.
$E^*(IAS)\!= \!13015\,(4)\,keV$ is the previously obtained
excitation energy of the IAS in $^{26}Si$ and 
$\Delta_{nH}$ is the mass excess difference between a neutron and
a hydrogen atom. $\Delta E_c$ is the Coulomb energy difference
between the IAS of $^{26}$Si and the ground state of $^{26}P$. It can be
deduced from the semi-empirical relation given in reference
\cite{[an86]}:

\begin{equation} \Delta E_c\,=\,1440.8*\left(\frac{\bar{Z}}
{A^{\frac{1}{3}}}\right)\:-\:1026.3
\end{equation}

Taking $\bar Z\!=\!14.5$ as a mean atomic number for the two
$A=26$ nuclei, the atomic mass excess of $^{26}P$ was deduced to
be equal to $11114\,(90)\,keV$. This value is in agreement 
with the mass prediction from Audi {\it et al.}~\cite{[au97]} 
of 10970(200)~$keV$. It leads to a $Q_{EC}$ value of $Q_{EC}\!=\!\Delta
(^{26}P)\!-\!\Delta (^{26}Si)$ = $18258\,(90)\,$ $keV$.

{\it Search for other charged-particle emission modes}: The one-proton separation
energy of $^{26}P$ is given by the relation $S_p(^{26}P)$ $\!=\!
\Delta (H)\!+\!\Delta (^{25}Si)\!-\!\Delta (^{26}P)$, where
$\Delta (H)$ and $\Delta (^{25}Si)$ are the atomic mass excesses
of hydrogen and $^{25}Si$ \cite{[au97]}. The deduced value
$S_p(^{26}P)\!=\!0\,(90)\,keV$ suggests that $^{26}P$ can hardly
be a direct proton emitter since the available energy for
such a disintegration would not exceed $100\,keV$.

In the same way, the relevance of $\beta$-delayed $\alpha$
emission can be discussed. Assuming that $\alpha$ particles would
be emitted by the IAS of $^{26}Si$, the corresponding
$Q_\alpha$ value is given by the following relation:

\begin{equation}
Q_\alpha\,=\,E^*(IAS) + \Delta (^{26}Si) - \Delta (^{22}Mg) - \Delta (^4He)
\end{equation}

\renewcommand{\arraystretch}{1.}
\begin{table*}
\begin{center}
\begin{tabular}{|r r r r r|r|r|} \hline
\multicolumn{5}{|c|} { C.M. proton energy ($keV$) from the decay to
$^{25}Al$ states }&\multicolumn{2}{|c|}{$^{26}Si$ proton} \\
\multicolumn{5}{|c|} {} &\multicolumn{2}{|c|}{emitting states}\\
\cline{6-7}
{\raisebox{1.ex}[0pt]{$\frac{5}{2}^+_1;\,0$}}&{\raisebox{1.ex}[0pt]{$\frac{1}{2}^+_1;\,452
$}} &{\raisebox{1.ex}[0pt]{$\frac{3}{2}^+_1;\,945 $}}
&{\raisebox{1.ex}[0pt]{$\frac{7}{2}^+_1;\,1612 $}}  &
{\raisebox{1.ex}[0pt]{$\frac{5}{2}^+_2;\,1790$}}  &Energy &B.R.
($\%$) \\ \cline{1-5}
$1\!:\!\,412\,(2)$      & & & & &$
5929\,(\:\:5)$ &$17.96\,(90)$\\ \hline $2\!:\!\,778\,(3)$ & & & &
&$6295\,(\:\:6)$ &$0.78\,(\:\:7)$\\
 \hline $3\!:\!\,866\,(2)$ & &
& & &$ 6384\,(\:\:5)$ &$ 1.71\,(15)$\\ \hline $4\!:\!1248\,(2)$ &
& & & &$ 6765\,(\:\:5)$ &$ 1.51\,(12)$\\ \hline $8\!:\!1983\,(2)$
& & & & &$ 7501\,(\:\:5)$ &$ 2.39\,(16)$\\ \hline
                           &$6\!:\!1638\,(3)$&                           &                           &                           &$ 7606\,(\:\:6)$ &$ 0.65\,(\:\:6)$\\  \hline
                           &                           &$5\!:\!1499\,(2)$&                           &                           &$ 7962\,(\:\:5)$ &$ 0.99\,(\:\:7)$\\  \hline
$13\!:\!2638\,(18)$        &                           & & & &$
8156\,(21)$ &$ 0.11\,(\:\:4)$\\ \hline $14\!:\!2732\,(4)$ &
&$7\!:\!1798\,(4)$& &                           &$ 8254\,(\:\:5)$
&$ 0.67\,(\:\:7)$\\ \hline
                           &$12\!:\!2593 (13)$&&&&$8563\,(17)$ & $0.27\,(\:\:6)$\\\hline
\multicolumn{3}{|r}{$16\!:\!2908 (11)$}&&&$9370\,(15)$ &$
0.06\,(\:\:5)$\\ \hline
                           $23\!:\!3920\,(5)$&&$17\!:\!2968\,(5)$&
$10\!:\!2288\,(3)$&$9\!:\!2139\,(4)$&$
9433\,(\:\:4)$&$3.54\,(20)$\\  \hline
                           &$20\!:\!3766\,(9)$&$19\!:\!3258\,(4)$&
                           &                           &$ 9725\,(\:\:7)$ &
                           $ 0.59\,(\:\:8)$\\  \hline
$26\!:\!4793\,(3)$        & &$21\!:\!3817\,(6)$& & &$10299\,(6)$
&$ 0.67\,(\:\:7)$\\  \hline
                           &                           &
                           &                           &$18\!:\!3097\,( 6)$&
                           $10405\,(\:\:5)$ &$ 0.31\,(\:\:6)$\\  \hline
                           &$25\!:\!4719\,(6)$&                           &
                           &                           &$10688\,(\:\:9)$ &
                           $ 0.24\,(\:\:4)$\\  \hline
                           &$27\!:\!4858\,(4)$&                           &
                           &                           &$10827\,(\:\:8)$ &
                           $ 0.44\,(\:\:5)$\\  \hline
$33\!:\!7494\,(4)$&$32\!:\!7039\,(5)$&$31\!:\!6551\,(4)$&$30\!:\!5893\,(4)$&
$29\!:\!5710\,(3)$&
$13015\,(4)$&$ 3.12\,(20)$\\
 \hline\hline \multicolumn{5}{|c|} {
C.M. two-proton energy ($keV$) from the decay to $^{24}Mg$
states}&\multicolumn{2}{|c|}{$^{26}Si$ two-proton  }
\\ \multicolumn{5}{|c|} {} & \multicolumn{2}{|c|}{emitting states}
\\\cline{6-7}
{\raisebox{1.ex}[0pt]{$\;\;0^+_1;\:\:\:\:0$}}
&{\raisebox{1.ex}[0pt]{$2^+_1;\:1369 $}} & & & & Energy& B.R.
($\%$)\\\cline{1-5} $28\!:\!5247\,(3)$ &$22\!:\!3879\,(3)$& & & &$13036\,(
4)$ &$ 2.16\,(24)$\\  \hline
\end{tabular}
\caption{Excitation energies and $\beta$ feeding of the
    proton-unbound excited states of $^{26}Si$. They are deduced from
    the data compiled in table \ref{tab:26pprot}. On the left-hand side, in the 
    top row and in the row last but one, we give the final states on which the
    one- or two-proton emission ends. Then we indicate
    the peak numbers according to figure~\ref{26pprot} and their 
    center of mass proton energy.
    On the right-hand side, we use this information to determine the 
    excitation energy of the emitting states in $^{26}Si$ as well 
    as the $\beta$-decay branching ratio for the feeding of 
    these states.}
\label{26pfeeding}
\end{center}
\end{table*}

Taking into account the atomic mass excesses given in 
reference \cite{[au97]} and the excitation energy of the IAS
measured in this work, the available energy in such a
$\beta$-delayed $\alpha$ decay would be equal to
$3842\,(15)\,keV$. The $\alpha$ transitions towards the ground
state and the first two excited states of $^{22}Mg$ are therefore
energetically possible and would lead to three
$\alpha$ groups at $3840$, $2600$ and $530\,keV$, the last two
being followed by $\gamma$ rays at $1250$ and $2060\,keV$. No
evidence for such $\gamma$ lines was observed and the proton
groups whose energy could match with an $\alpha$ decay to the
ground state of $^{22}Mg$ (transitions $21$ and $22$ in table
\ref{tab:26pprot} and \ref{26pfeeding}) were convincingly identified
as $^{26}P$ $\beta$-delayed one- and two-proton transitions.
In addition, these $\alpha$ transitions would be forbidden transitions.
Hence, we conclude that $\beta$-delayed one- and two-proton emission are
the only decay modes of the IAS in $^{26}Si$.

The excitation energy and the $\beta$ feeding of
proton emitting states in $^{26}Si$ deduced from the present
analysis are given in table \ref{26pfeeding}. The summed feeding
of $^{26}Si$ proton-unbound excited states is deduced to be equal
to $39\,(2)\,\%$.

%%%%%%%%%%%%%%%%%%%%%%%%%%%%%%%%%%%%%%%%%%%%%%%%%%%%%%%%%%%%%%%%
\subsubsection{$\beta$-delayed $\gamma$ decay}

\begin{figure}
\begin{center}
\resizebox{.48\textwidth}{!}{\includegraphics{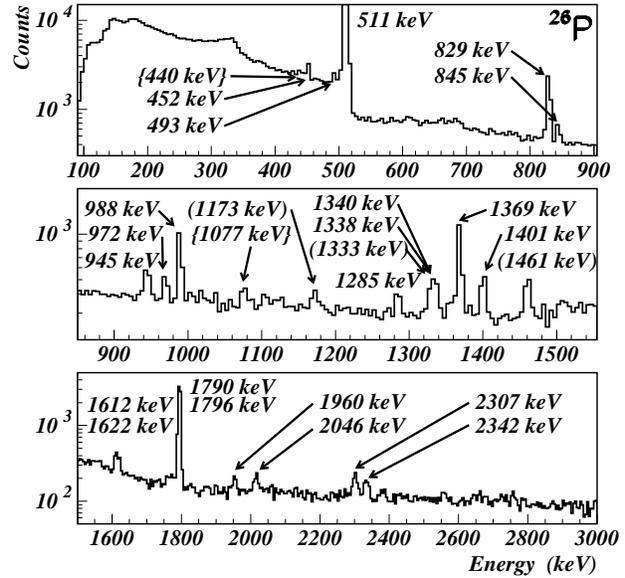}}
\caption[]{$\gamma$-decay spectrum of $^{26}P$: Background $\gamma$ lines from $^{60}$Co 
         and $^{40}$K are indicated within parenthesis. $\gamma$ rays within 
         brackets are due to the contaminating nuclei $^{23}$Mg and $^{24}$Al.
         All other $\gamma$ lines are related to the $\beta$ decay of $^{26}$P, 
         except the one at 1285 keV which could not be attributed.}
\label{26pgam}
\end{center}
\end{figure}

The $\gamma$-ray spectrum obtained during the $^{26}P$ setting is
shown in figure \ref{26pgam}. The six $\gamma$ lines at $972$
($1.27\,(54)\,\%$), $988$ ($5.2\,(11)\,\%$), $1796$
($52\,(11)\,\%$), $1960$ ($1.32\,(34)\,\%$), $2046$
($1.44\,(40)\,\%$) and $2342\,keV$ ($1.28\,(51)\,\%$) are
assigned to the $\beta$-delayed $\gamma$ decay of $^{26}P$
according to reference \cite{[en90]}. A new $\gamma$ ray at
$1400.5\,(5)\,keV$ ($2.82\,(69)\,\%$) was attributed to the
deexcitation of the $4184\,keV$ excited state of $^{26}Si$ to the
state at $2784\,keV$. The measured absolute intensities were
combined with previously measured relative $\gamma$-ray intensities \cite{[en90]}
to determine absolute $\beta$ feedings. The 
values obtained are reported in table \ref{26pfeeding2}.

The $\beta$ decay of the ground state of $^{26}Si$ is followed by
two $\gamma$ rays at $830$ and $1622\,keV$. The two lines at
$1341$ and $2307\,keV$ are due to the summing of the $511\,keV$
$\gamma$ rays (from the annihilation of the emitted positrons)
with the intense $\gamma$ rays at $830$ and $1796\,keV$.

The $\beta$-delayed one-proton and two-proton decays towards excited
states in $^{25}Al$ and $^{24}Mg$ give rise to the observation of
the $\gamma$ rays at $452$, $493$, $845$, $945$,
$1612$, $1790\,keV$ and $1369\,keV$. The absolute intensity of the
$1790\,keV$ $\gamma$ line, which forms a doublet with the $\gamma$
line at $1796\,keV$, was deduced from the previous determination
of the proton-group intensities to the related excited level in
$^{25}Al$. Some of these $\gamma$ lines are also due to the
presence of contaminating nuclei: $^{25}Si$ (see chapter 3.1), 
$^{24}Al$ ($\gamma$ rays at $1073$ and
$1369\,keV$) and $^{23}Mg$ ($\gamma$ ray at $440\,keV$). Background 
$\gamma$ lines are also visible in the spectrum at
$1461\,keV$ ($^{40}K$), and at $1173$ and $1333\,keV$ ($^{60}Co$).
The $\gamma$ ray at 1285~$keV$ could not be assigned.

According to tables \ref{26pfeeding} and \ref{26pfeeding2}, the summed
feeding of $^{26}Si$ proton-unbound  and proton-bound states
is equal to $39\,(2)\,\%$ and $54\,(12)\,\%$, respectively. The
spin of the ground state of the even-even nucleus $^{26}Si$ being
equal to $0^+$ \cite{[en90]}, this state is not expected to be fed 
significantly by a second forbidden $\beta$ decay of the $3^+$ ground
state of $^{26}P$. The summed feeding of the excited states of
$^{26}Si$ obtained in the present work is therefore equal to
$93\,(13)\,\%$. Taking into account the large uncertainty, the
result is in agreement with the expected value of $100\,\%$.
However, unidentified weak proton groups or $\gamma$ lines (see 
comparison to shell-model calculations below) may also contribute
to the missing strength.
In addition, it cannot be excluded that the $\beta$ feeding of the
$1796\,keV$ state was underestimated or that the $\gamma$ decay of
proton-bound states lying in the gap between $4184$ and $5929\,keV$ of
excitation energy (see decay scheme, figure~\ref{26ptrans}) was not observed.

%%%%%%%%%%%%%%%%%%%%%%%%%%%%%%%%%%%%%%%%%%%%%%%%%%%%%%%%%%%%%%%%
\subsubsection{Measurement of the half-life of $^{26}P$ }

\begin{figure}
\begin{center}
\resizebox{.48\textwidth}{!}{\includegraphics{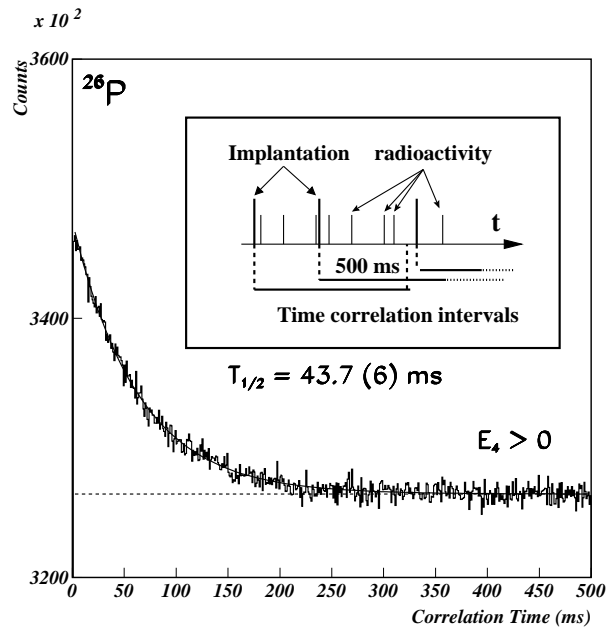}}
\caption[]{Determination of the $^{26}P$ half-life by
     means of timing correlations
     between implantation and radioactivity events within a time window
     of $500\,ms$. The inset shows the correlation of each implantation event 
     with each subsequent decay event within 500~$ms$.}
\label{26plife}
\end{center}
\end{figure}

The lifetime of $^{26}P$ was determined by means of a time
correlation procedure. The applied technique is schematically shown in the
inset of figure \ref{26plife}. It consists in measuring the time
difference between the implantation of $^{26}P$ ions, identified
by means of time-of-flight and energy-loss measurements, and the
observation of $\beta$ or $\beta$(2)p decay events.

Decay events that are
correlated to the selected implantation event follow an exponential decay
curve, whereas uncorrelated events due to the decay of
contaminant ions, due to $^{26}P$ daughter nuclei or due to $^{26}P$ implantations
other than the one considered for the correlation are randomly
distributed. The large time correlation window of $500\,ms$
enabled us to estimate accurately the contribution of uncorrelated
events to the decay curve. The half-life of $^{26}P$ was measured
to be $43.7\,(6)\,ms$, in agreement with the value given by 
Cable {\it et al.} \cite{[ca84]} of $20^{+35}_{-15}\,ms$.
We verified that, due to its relatively long half-life (2.21~s), 
the daughter decay of $^{26}$Si does not alter the fit result.

\renewcommand{\arraystretch}{1.1}
\begin{table}
\begin{tabular}{|c|c|r|} \hline
\multicolumn{1}{|c|}{$^{26}Si$ populated states}
&\multicolumn{1}{c|}{Excitation energy ($keV$)}
&\multicolumn{1}{c|}{ B.R. ($\%$)}
\\\hline   $2^+_1$  &$1795.9 \:(2)$&$44 \quad(12)$\\
 $2^+_2$&$2783.5 \:(4)$&$3.3
\:\:(20)$  \\
 $(\,3^+_1\,)$&$3756 \quad(2) $
&$2.68 \:(68)$    \\
 $(\,4^+_1\,)$&$3842 \quad(2) $ &$1.68 \:(47)$    \\
 $ 2^+$&$4138 \quad(1) $ &$1.78 \:(75)$    \\
 $(\,3^+_2\,)$&$4184 \quad(1) $ &$2.91
\:(71) $  \\ \hline
\end{tabular}
\caption{$\beta$-decay branching ratios towards proton-bound
excited states of $^{26}Si$.}\label{26pfeeding2}
\end{table}

%%%%%%%%%%%%%%%%%%%%%%%%%%%%%%%%%%%%%%%%%%%%%%%%%%%%%%%%%%%%%%%%
\subsubsection{$\beta$-decay scheme of $^{26}P$}

\begin{figure*}
\begin{center}
\resizebox{.8\textwidth}{!}{\includegraphics{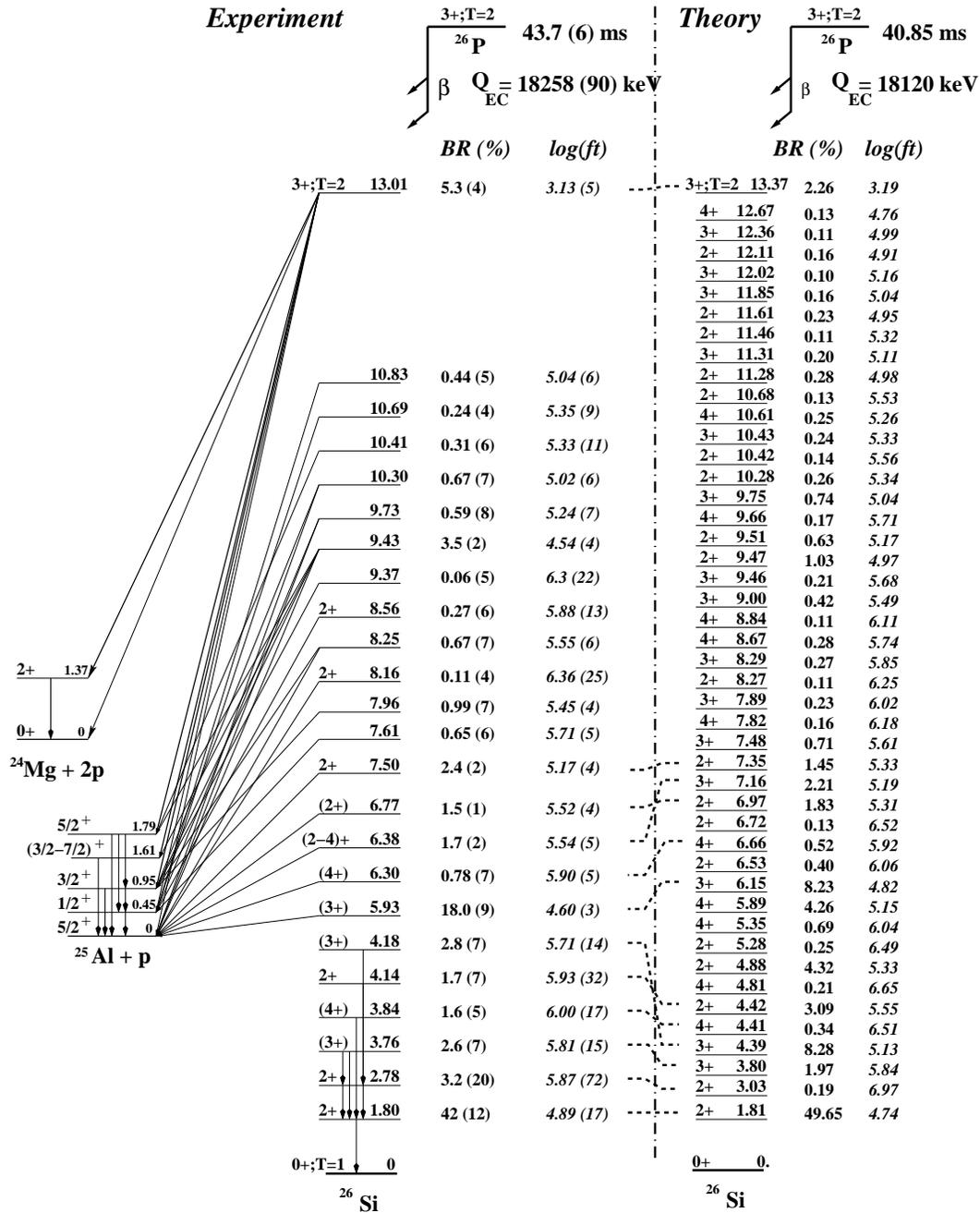}}
\caption[]{$^{26}P$ $\beta$-decay
            scheme as deduced from the data presented in this work.
            The dotted lines tentatively connect experimentally determined 
            levels to levels predicted by theory.}
\label{26ptrans}
\end{center}
\end{figure*}

The proposed $\beta$-decay scheme of $^{26}P$ is shown in figure
\ref{26ptrans}. The measured half-life as well as the $Q_{EC}$
value obtained experimentally are reported. The distribution of
$^{26}Si$ excited states appears to be well reproduced up
to an energy of $7\,MeV$ by the
shell-model calculations performed by Brown \cite{[bab]}. 

\begin{figure}
\begin{center}
\resizebox{.48\textwidth}{!}{\includegraphics{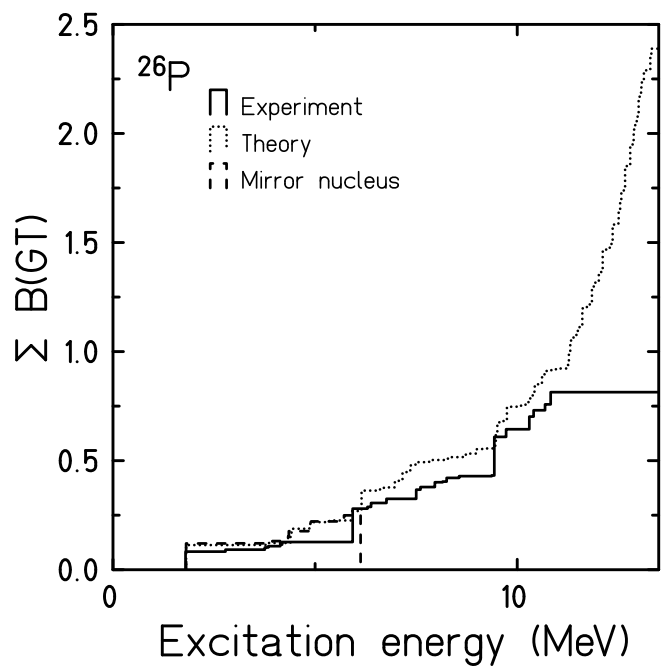}}
\caption[]{Summed Gamow-Teller strength distribution
        in the decay of $^{26}P$. The result of the present experiment 
        (relative error of about 20\%) is
        compared to shell-model calculations and to the B(GT) of the mirror
        decay of $^{26}$Na assuming isospin symmetry.}
\label{26bgt}
\end{center}
\end{figure}

As shown in figure \ref{26bgt}, the summed Gamow-Teller strength
distribution is also well reproduced up to an excitation energy of
more than $10\,MeV$. It can therefore be concluded that the
quenching of the Gamow-Teller strength in the $\beta$ decay of
$^{26}P$ is about $60\,\%$, as it was the case for $^{25}Si$.

At low excitation energy, the experimental Gamow-Teller strength
distribution is in disagreement with the one derived theoretically
between $4$ and $6\,MeV$. The discrepancy can be due to the
non-observation of populated excited states in this region, as it
was mentioned before. The summed $B(GT)$ strength
converges around $6\,MeV$ because of the high $\beta$-decay branching
ratio towards the excited state at $5.93\,MeV$. Hence, the noticed
discrepancy may simply originate from a different sharing of the 
$\beta$-decay strength between the two competing $3^+$ excited states at
$3.76$ and $5.93\,MeV$.

Once again, the uncertainty of the
$\beta$-decay branching ratios to low-energy excited
states is too large (see table~\ref{tab:asymm} below) to derive precise values for the
asymmetry parameter $\delta$ for the $(A\!=\!26,T\!=\!2)$ isospin
multiplet ($^{26}Na$,$^{26}Mg$,$^{26}Si$,$^{26}P$).

%%%%%%%%%%%%%%%%%%%%%%%%%%%%%%%%%%%%%%%%%%%%%%%%%%%%%%%%%%%%%%%%
%%%%%%%%%%%%%%%%%%%%%%%%%%%%%%%%%%%%%%%%%%%%%%%%%%%%%%%%%%%%%%%%
\subsection{Mirror asymmetry of mass A = 25, 26 nuclei}

The mirror asymmetry parameter $\delta$ is usually determined 
for the ground-state transitions as well as for those feeding
the low-lying excited states in the daughter nuclei. Higher-lying states
are normally fed with smaller branching ratios which yields larger
errors for the $\delta$ value. In addition, these states may decay
by proton emission for the proton-rich partner which usually reduces
the branching-ratio precission.

In the present experiment, however, the feeding of low-lying states and 
in turn also of the ground state (its branching ratio is determined
as the difference between 100\% and the observed branchings) is only 
poorly determined due to the large uncertainties of the $\gamma$-ray 
efficiency of our set-up. Nontheless, we give the asymmetry values
derived from the present work for the mass $A=25$ and $A=26$ nuclei 
in table~\ref{tab:asymm}.

\begin{table}[hht]
\begin{tabular}{|cccc|r|r|}
\hline\rule{0pt}{1.3em}
A   & $E^*(MeV)$ & $ J_i^\pi;T_i$ & $J_f^\pi;T_f$ & theory~\cite{[nadya]} & present work\\
[0.5em]\hline\rule{0pt}{1.3em}
25  & 0.00 & $\frac{5}{2}^+;\frac{3}{2}$ & $\frac{5}{2}^+;\frac{1}{2}$ &  1.11 &  0(40) \\
    & 0.95 &                             & $\frac{3}{2}^+;\frac{1}{2}$ & 12.39 &  0(20) \\
    & 1.61 &                             & $\frac{7}{2}^+;\frac{1}{2}$ & 11.23 & 30(40) \\
    & 2.67 &                             & $\frac{5}{2}^+;\frac{1}{2}$ & -5.58 & 48(11) \\
[0.5em]\hline\rule{0pt}{1.3em}
26  & 1.81 & $3^+;2$ & $2^+;1$ &  & 50(60) \\
    & 3.76 &         & $3^+;1$ &  & 10(40) \\
    & 4.14 &         & $2^+;1$ &  & 110(160) \\
    & 4.18 &         & $3^+;1$ &  & 110(70) \\
    & 5.93 &         & $3^+;1$ &  & -24(11) \\
[0.5em]\hline
\end{tabular}
\caption[]{Mirror asymmetries for the decay of $^{25}$Si and $^{26}$P
           and their mirror nuclei to low-lying states in the daughter
           nuclei. The excitation energies indicated in the second column
           are those of the $\beta^+$ daughter nuclei. Initial and final
           spin and isospin values are given in the following two columns.
           The experimental asymmetry results from our experimental values
           for the $ft$ values of the $\beta^+$ decay and data from
           the literature~\cite{[en90]}. The theoretical result is from 
           ref.~\cite{[nadya]} where we took the INC+WS value as one example.
           The last transition for each mirror pair stems from the measurement
           of a $\beta$-delayed proton branch for the proton-rich nucleus.}
\label{tab:asymm}
\end{table}

Experimentally, we reach the best precision for the highest-lying state
in each mirror couple where the $ft$ value for the proton-rich nucleus
comes from a $\beta$-delayed proton branch. In both cases, a significant
effect is observed. This result, however, is opposite in sign compared to
the theoretical value for the mass A=25 couple. For the other mirror transitions, 
no clear statement can be made due to the large experimental errors for the
decay of the proton-rich partner.

%%%%%%%%%%%%%%%%%%%%%%%%%%%%%%%%%%%%%%%%%%%%%%%%%%%%%%%%%%%%%%%%
%%%%%%%%%%%%%%%%%%%%%%%%%%%%%%%%%%%%%%%%%%%%%%%%%%%%%%%%%%%%%%%%
%%%%%%%%%%%%%%%%%%%%%%%%%%%%%%%%%%%%%%%%%%%%%%%%%%%%%%%%%%%%%%%%
\section{Conclusion and perspectives}

The $\beta$ decay of the neutron-deficient nuclei $^{25}Si$ and $^{26}P$
was studied at the LISE3 facility at GANIL.  $300$
and $60$ ions per second, respectively, were produced with contamination rates
of less than $1\,\%$ and of about $13\,\%$. The decay scheme of the two nuclei was
obtained, including for the first time the $\beta$-decay pattern
towards proton-bound states. It allowed us to measure the asymmetry parameter $\delta$ 
for the mirror states of the mass A=25 and A=26 nuclei. 
Unfortunately, the poor precision in the determination 
of the corresponding branching ratios gave rise to large uncertainties 
for these $\delta$ values. Comparison to shell-model
calculations based on the USD interaction and performed in the
full $sd$ shell by Brown \cite{[bab]} revealed two features:
the reliability of such models when they are applied to
mid-shell nuclei lying close to the proton drip-line, and the
about $60\,\%$ quenching of the
Gamow-Teller strength of the individual $\beta$ transitions.

The following properties were derived from the spectroscopic study
of these nuclei: i) The half-life of $^{26}P$ was measured to be
equal to $43.7\,(6)\,ms$. ii) Its proton separation energy as well as
the maximum available energy in its $\beta$ decay were determined
with a precision of $90\,keV$. iii) The $\beta$-delayed two-proton
emission of $^{26}P$ towards the ground state and the first
excited state of $^{24}Mg$ was observed. iv) More than thirty
one-proton groups were identified, five of them being emitted
from the isobaric analog state of $^{26}Si$.

Compared to previous studies with a helium-jet technique, 
the use of projectile fragmentation in conjunction with
a fragment separator has several advantages, which are that
i) the detection of the arrival of an ion allow its identification  
and gives a start signal for half-life measurements, ii) 
very short half-lives can be studied since the separation time is 
short (order 1 microsecond), iii) the selection process is
independent on chemistry.

Nonetheless, the spectroscopic studies presented here suffer from 
limitations that should be addressed in future experiments of 
the same type. Firstly, the implantation of ions inside a
silicon detector gives rise to a high proton-detection efficiency, 
however, due to the energy deposit of $\beta$ particles 
in the implantation detector, it is sometimes difficult to observe 
$\beta$-delayed protons with low intensity.

Secondly, concerning the $\gamma$-spectroscopy part, a
high efficiency is required in order to identify low intensity
$\gamma$ rays, and a high precision is needed for the more intense
transitions. A new detection set-up using segmented silicon
detectors and four Germanium clovers has therefore been
implemented and the decay properties of $^{21}Mg$, $^{25}Si$ and
their mirror nuclei were investigated recently at the GANIL
facility \cite{[gi03]}. This work should lead to the determination
of accurate asymmetry parameters $\delta$, which might help 
to understand the origin of isospin non-conserving
forces in nuclei.

%%%%%%%%%%%%%%%%%%%%%%%%%%%%%%%%%%%%%%%%%%%%%%%%
%%%%%%%%%%%%%%%%%%%%%%%%%%%%%%%%%%%%%%%%%%%%%%%%
\begin{acknowledgement}
The author would like to thank B.A. Brown for providing up-to-date
shell-model calculations and C. Volpe and N.A. Smirnova for
stimulating discussions about the mirror asymmetry question.
We would like to acknowledge the continous effort of the whole 
GANIL staff for ensuring a smooth running of the experiment. This work 
was supported in part by the Conseil R\'egional d'Aquitaine.

\end{acknowledgement}

%%%%%%%%%%%%%%%%%%%%%%%%%%%%%%%%%%%%%%%%%%%%%%%%
%%%%%%%%%%%%%%%%%%%%%%%%%%%%%%%%%%%%%%%%%%%%%%%%


\end{document}